# Linear Accelerators

*M. Vretenar*
CERN, Geneva, Switzerland

**Abstract**
The main features of radio-frequency linear accelerators are introduced, reviewing the different types of accelerating structures and presenting the main characteristics aspects of linac beam dynamics.

## 1 Introduction

In a *linear accelerator* (linac) charged particles acquire energy moving on a linear path; their characteristic feature is that particles pass only once through each of the accelerating structures[1]. In the following, we limit our analysis to radio frequency (RF) linacs where the acceleration is provided by time-varying electric fields, leaving out of our treatment *electrostatic linacs* based on DC fields and used only at low beam energies.

A linac will receive the particle beam coming out of an ion source, bunch it at a given RF frequency, and then accelerate it up to the required final energy. In general, linacs are *pulsed* accelerators: the beam is generated by the source and then delivered to the users in pulses of a duration between a few microseconds and a few milliseconds at a given repetition frequency. Linacs injecting into synchrotrons operate at the low repetition frequency imposed by the rise time of the synchrotron magnets, while stand-alone linacs can easily go to higher repetition frequencies often synchronised with the mains (60 Hz in the US, 50 Hz in Europe). Electron linacs usually operate with short pulse durations at high repetition frequencies, up to several hundreds of hertz. The product of pulse length and repetition frequency is the *duty cycle*; if the linac accelerates the beam continuously the duty cycle is 100% and the linac is said to operate in continuous wave (CW) mode.

Together with the *kinetic energy E* of the particles coming out of a linac, its most important parameter is the beam current $I$, defined as the *average current during the beam pulse*. The current $I$ is different from the average current, which is $I$ times the duty cycle; it can also be different from the *bunch current*, the average current during an RF period populated by particles, in the case when some of the RF periods ('buckets') are empty. An important parameter for proton linacs is the *beam power*, the electrical power transferred to the particle beam during the acceleration process, which is

$$P \text{ [W]} = E \text{ [eV]} \times I \text{ [A]} \times \text{duty cycle} . \tag{1}$$

The fact that the beam only goes once through each accelerating structure allows optimisation of the design of each structure for a specific particle velocity. In this way proton and heavy ion linacs can provide good acceleration efficiencies in the energy range where the velocity increases with energy, at the price of some mechanical complexity. Figure 1 shows how, for protons and electrons, the square of the particle velocity increases with the kinetic energy. In the case of protons, we can identify two regions, a 'Newton' region where the velocity rapidly increases with energy, and an 'Einstein' region where velocity tends to saturate towards the speed of light. When protons become relativistic, it is usually convenient to transfer the beam to a synchrotron where, after paying the entry price of the magnet system, the beam can be accelerated many times through the accelerating elements. For electrons that are almost immediately relativistic, linear accelerators are used up to high energies: the linac accelerating structures have a simple standard design and provide a high efficiency over the full energy range, while energy loss due to synchrotron radiation penalizes circular accelerators.

---

[1] It should be mentioned that linear accelerators do not need to be linear: some heavy ion linacs, for example, incorporate 90° bends to reduce the footprint of the machine. In a similar way, some electron linac designs include 180° bends, sometimes even passing the beam two or more times through the same accelerating structures; these are 'recirculating linacs'.

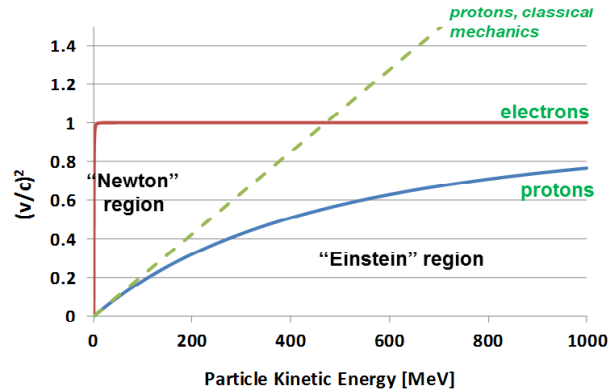

**Fig. 1:** Relativistic velocity squared as a function of kinetic energy for protons and electrons

Proton linacs are generally used at low energy and as injectors for synchrotrons, for their ability to efficiently accelerate beams of increasing velocity; nevertheless, recent years have seen a notable trend towards the design and construction of proton linacs at increasingly higher energies, well into the GeV range. The motivation is an increasing request for intense secondary particle beams (mainly neutrons, but also pions to generate neutrino beams and radioactive ions) produced by a high-intensity proton beam impinging on a target. In the production and acceleration of intense proton beams, linacs have a clear advantage with respect to synchrotrons because the higher repetition frequencies achievable with linacs permit distribution of the charges over more pulses and bunches, thus reducing space charge effects, and because instabilities and collective effects limiting beam intensity in synchrotrons play only a minor role in linacs, where the beam goes only once through each period. Additionally, even for linacs used as injectors to synchrotrons, selecting a high linac energy allows the accumulation of more current into the synchrotron. This is related to the fact that injection into the synchrotron takes place over many turns, accumulating bunch trains from the linac to populate the synchrotron acceptance. The final limit to the number of turns and to the total intensity comes from space charge-induced tune shift at injection into the synchrotron, an effect that decreases with energy.

## 2   Coupled cells, synchronicity and energy flow

An RF linear accelerator is made of a sequence of accelerating elements. In order to accelerate, each element has to generate an electric field in the region crossed by the beam, usually consisting of a 'gap' between two metallic plates (Fig. 2). The field in the gap oscillates with time at a given RF frequency; because of its time dependence, the electric field must be associated with a magnetic field, i.e. the electromagnetic energy associated with the fields must oscillate between a region of high electric field (the gap) and a surrounding region of high magnetic field. Each gap is therefore part of a 'RF resonator', a 'resonant cavity' that encloses a region of space where the electromagnetic energy is confined and oscillates between electric and magnetic field at a frequency defined by the cavity geometry. The cross-section of the simplest RF cavity and its associated gap is shown in Fig. 2; this is the so-called pill-box cavity, a cylindrical cavity with rotational symmetry around the beam axis. While the electric field is concentrated in the central gap region, the magnetic field turns around the gap. In Fig. 2 are reported the basic relations for the electric field in the gap and for the energy gained by a particle with electric charge $e$ crossing the gap at a phase $\phi$ relative to the crest of the E-field sinusoidal waveform. Here $V_0$ is the integral of the electric field over the gap length, while $T$ is the transit-time factor and represents the fraction of the field on the axis that can actually be seen by a particle travelling at a given velocity; it is always lower than 1.

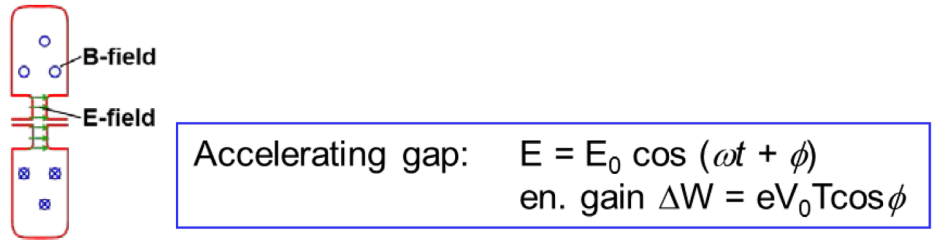

Fig. 2: Basic accelerating cell

The simplest linac structure is a sequence of accelerating elements like the pill-box shown in Fig. 2; this is shown in Fig. 3 where we assume that all RF cavities are identical and resonating at the same (angular) frequency $\omega$.

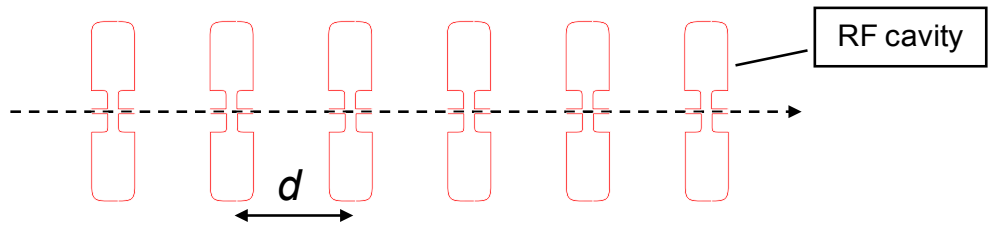

Fig. 3: Sequence of accelerating gaps and cavities

What are the conditions for a particle to be accelerated through this sequence of accelerating elements? To gain energy, particles have to cross each gap at a time when the electric field is positive ($\cos\phi > 0$). This is a simple condition to be fulfilled for one particle and one gap, but for a large number of particles crossing a large number of gaps and increasing their energy and their velocity this corresponds to two very specific conditions:

i) the particles must already be grouped together (i.e. the beam must be 'bunched') when they enter the sequence of gaps; time separation between bunches must be a multiple of the RF period $T = 2\pi/\omega$;

ii) the distances between the gaps and their relative RF phases must be correlated.

This second condition leads to an important relation: the electric field on the gap $i$ is
$$E_i = E_{0i} \cos(\omega t + \varphi_i) \tag{2}$$

with $\varphi_i$ the phase of the $i$th cavity with respect to a reference RF phase. To maximize acceleration, the beam has to cross each gap at a phase $\varphi_i$ on or very close to the crest of the wave ($\varphi_i = 0$ in the linac convention); moreover, it must have a short length in time and phase, i.e. it must be 'bunched'. During the time that the particles need to go from one cavity to the next the phase has changed by an amount $\Delta\varphi = \omega\tau$ with $\tau$ the time to cross a distance $d$; for a particle of relativistic velocity $\beta = v/c$ the change in phase will be
$$\Delta\Phi = \omega\tau = \omega\frac{d}{\beta c} = 2\pi\frac{d}{\beta\lambda} \tag{3}$$

where $d$ is the distance between gaps and $\lambda$ is the RF wavelength. This means that
$$\frac{\Delta\Phi}{d} = \frac{2\pi}{\beta\lambda} \tag{4}$$

or, for the acceleration to take place, the *distance* and the *phase difference* between two gaps in the sequence must be correlated, their ratio being proportional to $\beta\lambda$. At every gap crossing the particle will gain some energy and its velocity will increase: in a non-relativistic regime this means that either the relative phase $\Delta\Phi$ or the distance $d$ has to change during acceleration. In other terms, in a linear accelerator we need either to progressively increase the distance between cavities or to progressively decrease their RF phase (relative to a common reference) to keep synchronicity between the particle beam and the accelerating wave.

This requirement corresponds to two well-defined types of linear accelerators:

i) 'Individual-cavity' linacs (Fig. 4(a)), where the *distance* between cavities is fixed, and the phase of each cavity is individually adjusted to take into account the increase in beam velocity; to control the phase, each cavity has to be connected to an individual RF amplifier. This scheme has the advantage of maximum flexibility, being able to accelerate different ions or different charge states at different input and output energies; it just requires the entry of a different set of cavity phases for each ion. Its main disadvantage is the high cost, which is related to the number of individual RF cavities and amplifiers.

ii) 'Coupled-cell cavity' linacs (Fig. 4(b)), where the *phase* at each cavity/gap is fixed, and the distance changes accordingly with the beam velocity. When the difference in phase between the gaps is precisely defined, it is convenient to couple together several gaps into a common RF resonator and connect them to a single RF power source, reducing significantly the cost of the RF system. How to couple together gaps and resonators defining the phase difference between adjacent gaps will be discussed below. It should be noted that this scheme is cost-effective but not at all flexible: the physical distance between gaps is defined for a fixed increase in energy, i.e. for a given particle, a given energy range and a given acceleration gradient $E_0$.

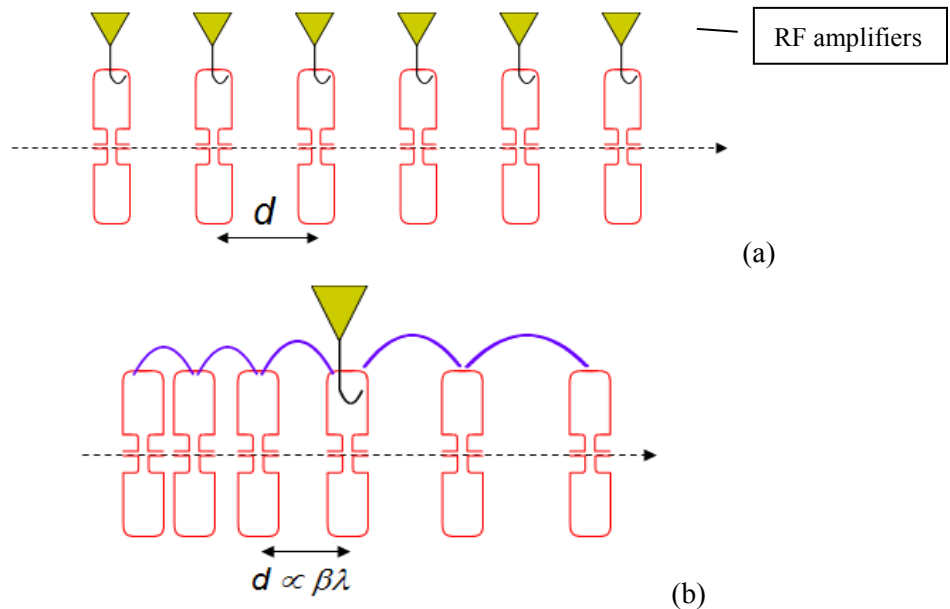

**Fig. 4**: (a) Single-cavity linac; (b) coupled-cell cavity linac

The reality of linac structures is not so rigidly defined and most of the commonly used linacs tend to compromise between these two extreme configurations. In particular, for high energies and high $\beta$'s, commonly used linac structures are made of sequences of equally spaced gaps. The reason becomes clear if we calculate the phase error, i.e. the difference between the nominal phase (the one at which the centre of the beam bunch should have crossed the gap) and the actual phase seen by a particle in the second gap of a sequence designed for a relative velocity $\beta$ when the beam has instead a velocity $\beta + \Delta\beta$. The phase error, usually called 'phase slippage' is

$$\Delta\varphi = \omega\Delta t = \pi\frac{\Delta\beta}{\beta} = \pi\frac{1}{\gamma(\gamma-1)}\frac{\Delta W}{W} \tag{5}$$

where $\Delta t$ *is* the error in time, $\beta$ and $\gamma$ the usual relativistic factors, and $W$ the kinetic energy.

The error is large at low energy and can bring the particles completely outside of the range of the accelerating phases. At high energies the factor $\gamma$ starts to increase rapidly and the error becomes increasingly small. For example, at 500 MeV and for an energy gain of 4 MeV/cell the phase error is only $\pi/100$ or about 1.8°, only a fraction of the usual phase length of a bunch. When the beam progresses inside a structure made of identical cells, the errors will add up and the beam phase will 'slip' from cell to cell; as soon as it remains close enough to the crest of the sinusoidal wave the loss in acceleration efficiency will be small while the mechanical simplification of having sequences of identical cells will considerably decrease the cost of the linac. For this reason, for high energies multi-cell linac structures are commonly used, such as the superconducting four-cell cavity shown in Fig. 5.

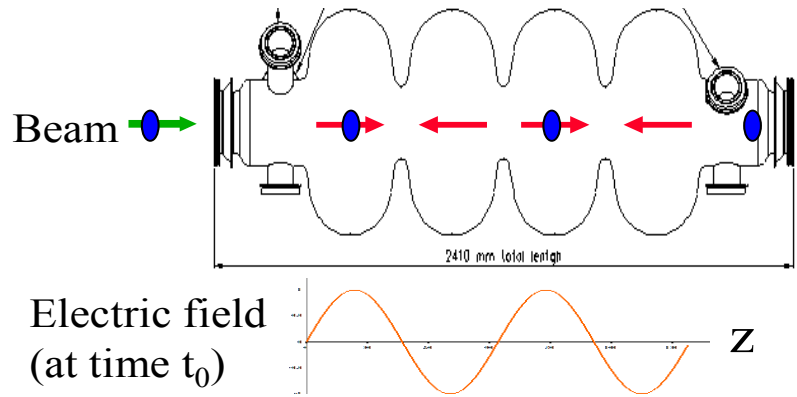

**Fig. 5:** Multi-cell superconducting cavity

The four 'gaps' of this cavity are coupled together in a single resonator. At a given time, the longitudinal electric field on the axis will have the profile shown in the figure and indicated by the arrows in the gaps: this corresponds to a constant phase difference $\Delta\Phi = 180°$ between adjacent gaps. For a particle to be accelerated, the relation between phase and gap distance must hold, with $\Delta\Phi = \pi$; solving for $d$, we find that for this particular system the condition for acceleration is

$$d = \frac{\beta\lambda}{2}, \tag{6}$$

i.e. the distance between gaps must be equal to $\beta\lambda/2$. This relation is valid only for one $\beta$, which can be the one at the centre of the cavity. Travelling inside this cavity, the beam will accumulate a phase slippage and the energy gained per cell will be lower than that given by the expression in Fig. 2. In a similar way, in a long sequence of structures with identical cell length there will be only one cell with length matched to the particle $\beta$; in all the others, acceleration will be lower. At energies above a few hundreds of MeV, however, the loss in acceleration is relatively small and is compensated for by the lower cost of series production of the accelerating structures. For this reason, a high-energy proton linac is always made of sections of identical accelerating cavities, the number of sections and the transition energies being the result of a careful cost optimisation.

Before going into the study of the different types of accelerating structures, it is important to consider that a linac is much more complex than a simple sequence of RF gaps. First of all, the sequence of gaps, usually grouped inside a cavity, has to be preceded by an ion source and by a bunching system (bunching will be discussed in Section 12). Then, the gaps have to be spaced by some focusing elements, usually quadrupoles, required to keep together the particles that constitute the bunch. The RF cavity needs to be fed by a RF amplifier inserted into a feedback loop and the system requires some ancillaries to work: a vacuum system, a magnet powering system, and a water-cooling system to evacuate the excess power coming from the RF system. A basic block diagram of a proton linac accelerating structure is presented in Fig. 6.

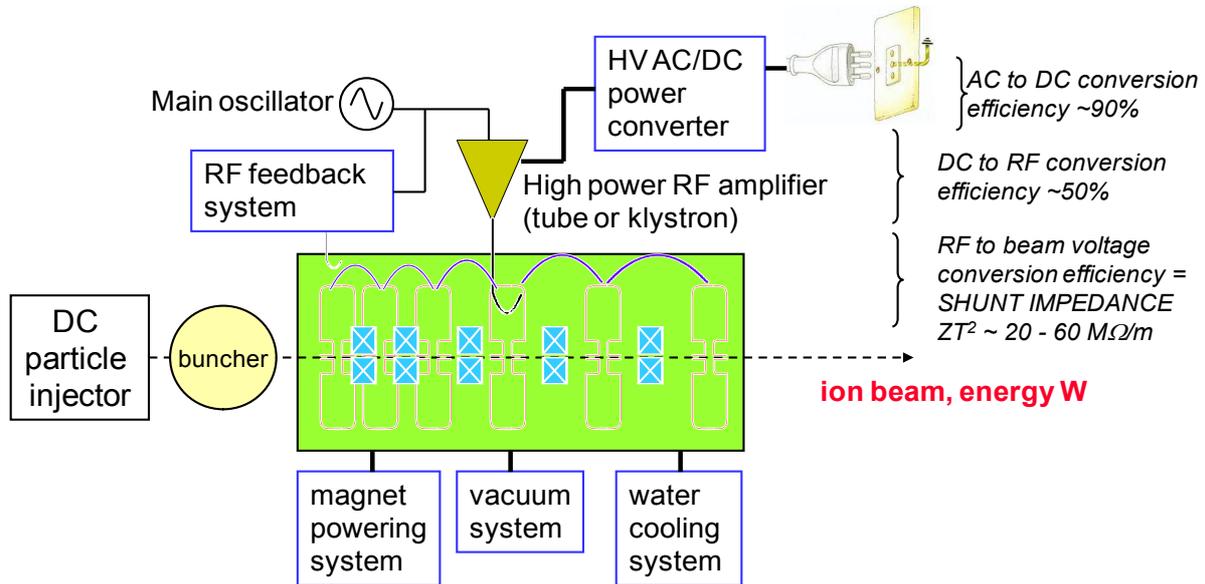

**Fig. 6**: Scheme of a linac accelerating system

The schematic representation in Fig. 6 shows how the main goal of a linac accelerating system is to transform the electrical energy coming from the grid into kinetic energy of the particle beam. Looking at the system from the point of view of power (energy per unit time), the accelerator converts a fraction of the power from the grid into the 'beam power' introduced in Section 1. The efficiency of this transformation is usually quite low, and a large fraction of the input energy is dissipated in heat released into the surrounding environment and/or removed by the cooling system. More precisely, the energy transformation takes place in three different steps, each one characterized by a different technology and a different efficiency:

i) The conversion of the AC power from the grid (alternating, low voltage, high current) into DC power (continuous, high voltage, low current) takes place in a power converter. Pulsed power converters for linacs are usually called 'modulators'; their efficiency is usually high, of the order of 90%. In case of pulsed operation the modulator is equipped with an energy storage system, usually a capacitor bank, to provide a constant load to the electricity network.

ii) The following conversion of DC power into RF power (high frequency, high voltage, low current) takes place in a RF active element: RF tube, klystron, transistor, etc. RF conversion efficiency depends on the specific device and on its class of operation. Typical RF efficiencies are in the range 50–60%.

iii) The final conversion of RF power into power given to the particle beam takes place in the gap of the accelerating cavity; its efficiency is proportional to the shunt impedance of the cavity, which represents the efficiency of the gap in converting RF power into the voltage available for a beam crossing the cavity at a given velocity (see Section 6).

## 3   Accelerating structures for linacs

Coupled-cell cavities are the most widely used linac accelerating structures. To couple the elements of a chain of single-gap resonators (these will be referred to below as the 'cells' of our system) we need to allow some energy to flow from one cell to the next, via an aperture that permits the leaking of some field (electric or magnetic) into the adjacent resonator. There will be two different types of coupling, depending on whether the opening connects regions of high magnetic field ('magnetic coupling') or regions of high electric field ('electric coupling'). The simplest magnetic coupling is obtained by opening a slot on the outer contour of the cell, whereas an electric coupling can be obtained by enlarging the beam hole until some electric field lines couple from one cell to the next. Once the cells are coupled, the relative phase in each gap is no longer free, but will depend on the parameters of the coupled system. To found the conditions for acceleration, we need to find the possible relations between the phases of the individual gaps.

The simplest way to analyse the behaviour of a chain of coupled oscillators is to consider their equivalent circuits (Fig. 7) [1].

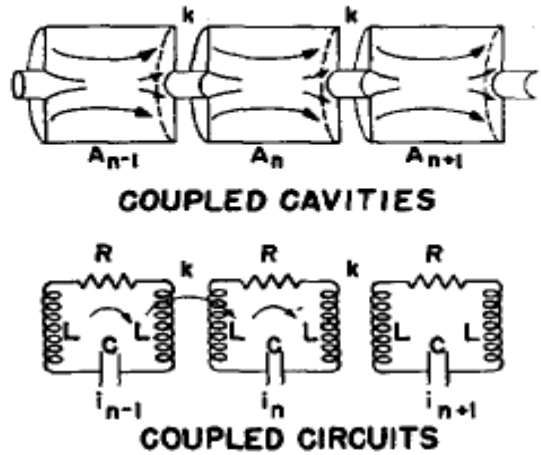

**Fig. 7**: From coupled cavities to coupled resonant electrical circuits (from Ref. [1])

Each coupled cavity can be represented by a standard *RLC* resonant circuit; in Fig. 7, for convenience, the inductance in each circuit is split into two separated inductances *L*. The advantage of this representation is that we can describe the magnetic coupling between adjacent cells as a mutual inductance *M* between two inductances *L*, which is related to an electrical coupling factor *k* by the usual relation $M = kL$. For the series resonant circuits in Fig. 7, the behaviour of each cell is described by its circulating current $I_i$. The equation for the *i*th circuit can be written taking the sum of the voltages across the different elements of the circuit (Kirchhoff's law) to be equal to zero, considering for simplicity a lossless system where $R = 0$

$$I_i \left(2j\omega L + \frac{1}{j\omega C}\right) + j\omega k L (I_{i-1} + I_{i+1}) = 0 \ . \tag{7}$$

Dividing both terms of this equation by $2j\omega L$, it can be written as

$$X_i \left(1 - \frac{\omega_0^2}{\omega^2}\right) + \frac{k}{2}(X_{i-1} + X_{i+1}) = 0 \ . \tag{8}$$

This equation relates general excitation terms of the form $X_i = I_i/2j\omega L$, proportional to the square root of the energy stored in the cell *i*, with the coupling factor *k* and with a standard resonance term $(1 - \omega_0^2/\omega^2)$. We consider that all cells are identical, i.e. that they have the same resonance frequency $\omega_0^2 = 1/(2LC)$. If our system is composed of $N + 1$ cells, taking $i = 0, 1,..., N$, we can write a system of $N + 1$ equations with $N + 1$ unknowns $X_i$ that can be expressed in matrix form

$$\begin{bmatrix} 1-\frac{\omega_0^2}{\omega^2} & \frac{k}{2} & 0 & & 0 \\ \frac{k}{2} & 1-\frac{\omega_0^2}{\omega^2} & \frac{k}{2} & \cdots & 0 \\ 0 & \frac{k}{2} & 1-\frac{\omega_0^2}{\omega^2} & & 0 \\ & \vdots & & \ddots & \vdots \\ 0 & 0 & 0 & \cdots & 1-\frac{\omega_0^2}{\omega^2} \end{bmatrix} \begin{vmatrix} X_0 \\ X_1 \\ \cdots \\ X_N \end{vmatrix} = 0 \qquad (9)$$

or

$$MX = 0 \ . \qquad (10)$$

The matrix $M$ has elements different from zero only on three diagonals: the main one composed of resonance terms and the two adjacent containing the coupling terms. The matrix is perfectly symmetrical because we have introduced an additional simplification assuming that the chain of resonators is terminated at both ends with 'half cells', with only one coupling $k$ but with half the inductance and twice the capacitance of a standard cell to keep the same resonant frequency. This corresponds to the physical case where the end resonators are terminated by a conducting wall passing through the centre of the gap, i.e. they are exactly one half of a standard cell. The advantage of this approach is that the matrix and the relative solutions are symmetric and lead to a simple analytical result. In a real case the chain of resonators will be terminated with full cells that need to be tuned to a slightly different frequency to symmetrise the system.

The matrix equation above represents a standard eigenvalue problem, which has solutions only for those $\omega$ giving

$$\det M = 0 \ . \qquad (11)$$

The eigenvalue equation $\det M = 0$ is an equation of $(N+1)$th order in $\omega$. Its $N+1$ solutions $\omega_q$ are the eigenvalues of the problem, which are the resonance modes of the coupled system.

The important information that we find here is that while the individual resonators oscillate only at the frequency $\omega_0$, the coupled system can oscillate at $N+1$ frequencies $\omega_q$ with $q = 0, 1,..., N$ the index of the mode. To each frequency corresponds a solution in the form of a set of $[X_i]_q$, which is the corresponding eigenvector.

Coupling together a number of identical resonating cells their resonance frequency will open up into a band of permitted frequencies, each one related to a particular distribution of the fields and of the energy between the cells. We can observe that this is a general feature of coupled systems in physics, as for example in the case of wave propagation within a crystal, a lattice of atoms or molecules bound by elastic forces. It is important to notice that the number of modes is always equal to the number of cells in the system.

In the case of the symmetric matrix $M$ we can find an analytical expression for the eigenvalues (mode frequencies)

$$\omega_q^2 = \frac{\omega_0^2}{1+k\cos\frac{\pi q}{N}} \qquad q = 0, ..., N \qquad (12)$$

or, for $k \ll 1$, which is the case for the coupled structures commonly used in linacs ($k \sim 1\text{–}5\%$)

$$\omega_q \approx \omega_0 \left(1 - \frac{1}{2} k \cos\frac{\pi q}{N}\right) \qquad q = 0, ..., N \ . \qquad (13)$$

The corresponding eigenvectors representing the field distribution between cells for each mode are

$$X_i^{(q)} = (\text{const}) \cos\frac{\pi q i}{N} e^{j\omega_q t} \qquad q = 0, ..., N \ . \qquad (14)$$

Equation (12) is particularly interesting, because it indicates that each mode $q$ is identified by a 'phase'

$$\Phi_q = \frac{\pi q}{N}. \tag{15}$$

The first mode, $q = 0$, has $\Phi = 0$ and frequency $\omega_{q=0} = \frac{\omega_0}{\sqrt{1+k}}$. The last mode, $q = N$, will have $\Phi = \pi$ and frequency $\omega_{q=N} = \frac{\omega_0}{\sqrt{1-k}}$. If we identify each mode by the value of $\Phi_q$ in radians the first can be called the '0' mode and the last the '$\pi$' mode. All other modes will have frequencies between the 0 and $\pi$ mode frequencies.

For $k \ll 1$ the difference between $\pi$ and 0 mode frequencies is

$$\Delta\omega = \omega_{q=N} - \omega_{q=0} = \omega_0 \left(\frac{1}{\sqrt{1-k}} - \frac{1}{\sqrt{1+k}}\right) \approx \omega_0 k, \tag{16}$$

i.e. the 'bandwidth' of the coupled system, the range covered by the coupled system frequencies, is proportional to the coupling factor $k$.

Plotting the frequencies given by Eq. (12) as a function of the phase $\Phi$, we obtain curves like that shown in Fig. 8, which corresponds to the case of five cells and five modes. This is a typical 'dispersion curve', relating the frequencies of our system with a propagation constant, in this case represented by the phase $\Phi$. The permitted frequencies lie on a cosine-like curve, and the modes are represented by points equally spaced in phase. The more cells in the system, the more modes we will have on the curve, until the limit of the continuous: for an infinite number of cells, all of the modes on the curve are allowed. The frequency separation between the modes depends on the coupling: the higher the coupling, the more separated in frequency will be the modes.

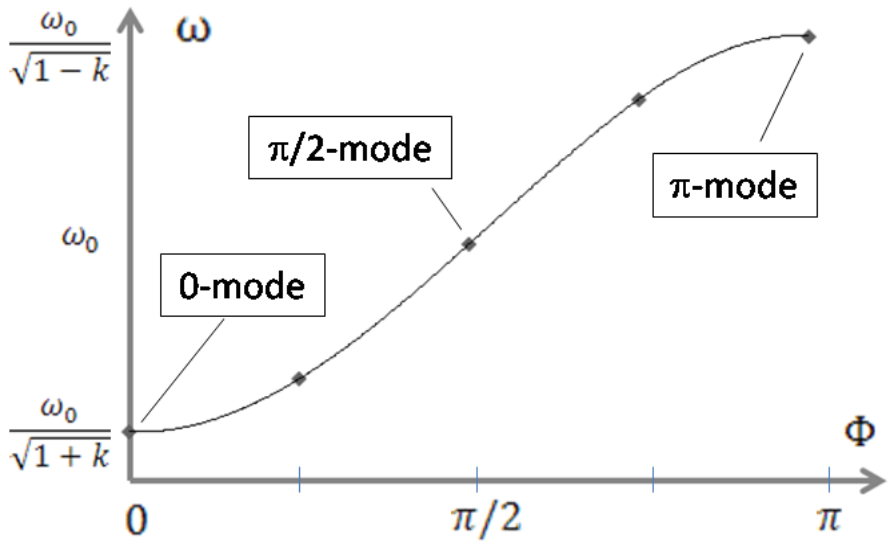

**Fig. 8**: Dispersion relation for a five-cell coupled resonator chain

The field distribution in the cells is defined by Eq. (14). For a given mode $q$, the fields will oscillate in each cell at the frequency $\omega_q$, and the amplitude of the oscillation will depend on the position of the cell in the chain. The distribution of maximum field amplitudes along the chain follows a cosine-like function with argument ($\Phi_q i$), i.e. the product of the phase $\Phi_q$ times the cell number $i$. It is now clear that $\Phi_q$ represents the *phase difference between adjacent cells* in the coupled system. We can now represent in Fig. 9 the field distribution between the cells in the chain for the main modes, for example for a seven-cell system with $N = 6$.

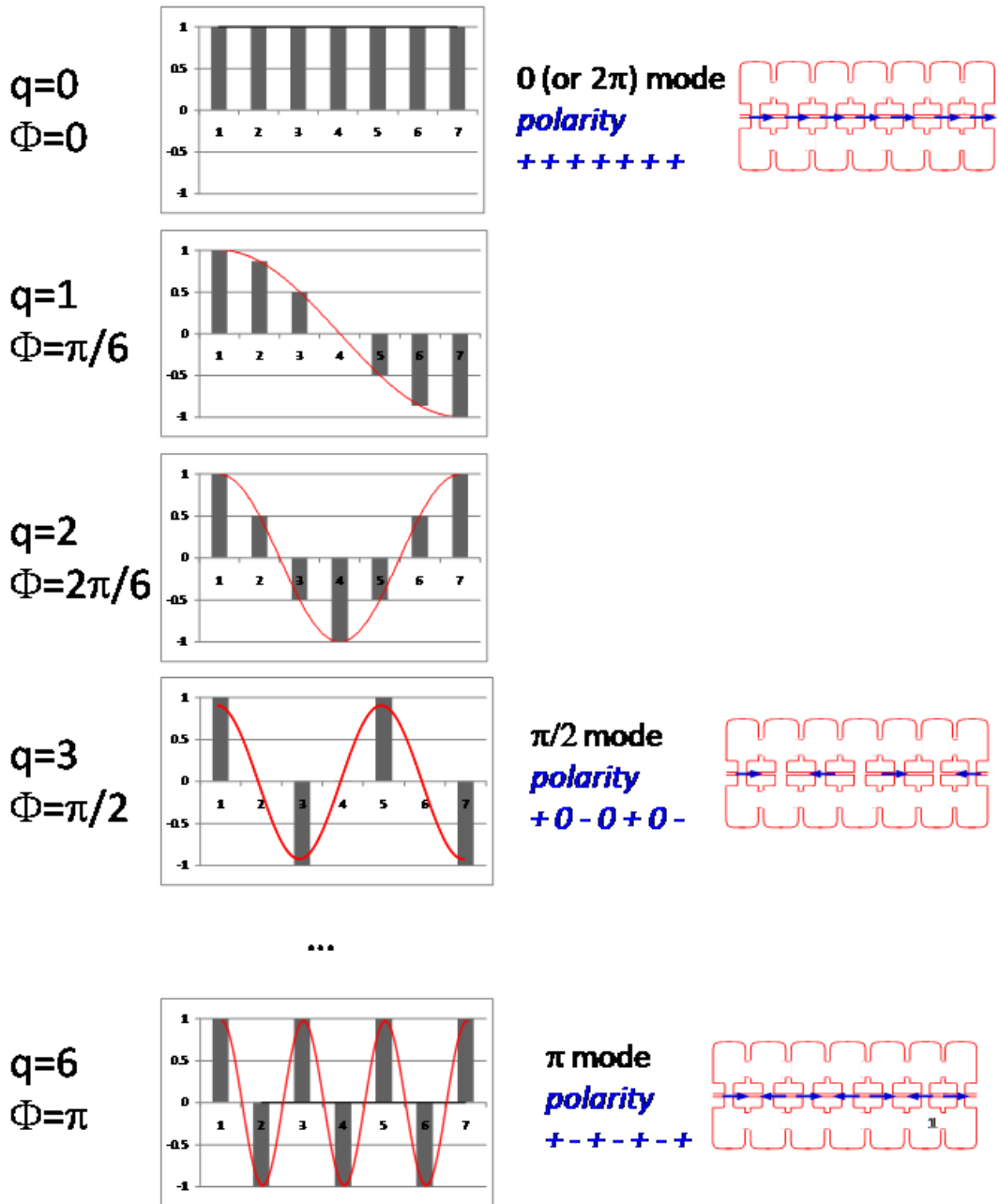

**Fig. 9**: Distribution of the fields in the cells of a seven-cell system and polarity of the electric field in the gaps for the modes used for particle acceleration

We can immediately observe that our solutions are standard 'standing-wave' modes, as in the case of similar physical systems. For example, a vibrating string composed of elastically bound molecules and fixed at both ends will follow the same mathematical description. It must be observed that Eq. (14) is made of two terms, a position term and a time-dependent term. In Fig. 9 we have plotted only the position term, assuming that the time-dependent term is 1 corresponding to time $t = 0$. Because the fields are oscillating with angular frequency $\omega$, the fields in each cell will periodically oscillate between the maximum represented in Fig. 9 and a minimum corresponding to fields with identical absolute value but reversed sign, reached at a time corresponding to $\omega t = \pi$. At $\omega t = \pi/2$ the field will be zero everywhere.

We have seen that as many modes as cells can be excited in a linac coupled-cell structure, each mode being characterized by a frequency and by a particular field distribution between cells, but how can these modes be used for the acceleration of particles? By selecting the frequency of the RF generator it is possible to precisely define the mode that is excited in the structure; we have now to see how its field distribution interacts with a charged particle travelling with velocity $\beta/c$.

From Eq. (12), the electric field in the cell $i$ can be written as the product of two cosine functions, one containing the space dependence and the other the time dependence. For a particle travelling with velocity $\beta/c$ between cells at distance $d$

$$E_{i,q} = E_{0,q} \cos(\Phi_q i) \cos\left(\frac{2\pi d}{\beta\lambda} i\right) \quad . \tag{17}$$

Maximum acceleration will correspond to the case where the product of the two cosine functions is 1. This corresponds to two particular situations.

i) $\Phi = 0$ (or $\Phi = 2\pi$) and $\cos(2\pi di/\beta\lambda) = 1$; this condition implies $d = \beta\lambda$. This is the 'zero mode' of acceleration. At a given time, the electric field is the same in all cells and during the time that a particle takes to go from one cell to the next the field makes a complete oscillation ($2\pi$ or 360°).

ii) $\Phi = \pi$ (or $\cos(\Phi i) = \pm 1$) and $\cos(2\pi di/\beta\lambda) = \pm 1$; this condition implies that the distance between cells is $d = \beta\lambda/2$. This is the '$\pi$ mode' of acceleration. At a given time, the electric field alternates in sign between adjacent cells, and during the time that a particle takes to go from one cell to the next the field makes a half oscillation ($\pi$ or 180°) and the particle again experiences an accelerating field.

The conclusion is that only the modes 0 and $\pi$ can be used for efficient particle acceleration. Once the mode of acceleration has been selected, the RF structure designer will define the geometrical dimensions of the structure in such a way that the 0 or $\pi$ mode will be excited at exactly the frequency of the RF generator.

An exception is the $\pi/2$ mode, which has $\cos 2\Phi_q = 0$. This can still be used for acceleration with $d = \beta\lambda/4$, but the acceleration is not very efficient, the field being present only in half of the cells. As we will see in the following, however, the $\pi/2$ mode presents the advantage of higher stability against deviations in the individual cell frequencies; this justifies its use for some specific accelerating structures.

## 4  Zero-mode structures: the drift tube linac

The first and most important structure operating in the 0-mode is the Drift Tube Linac (DTL), also called the Alvarez linac after the name of its inventor. It can be considered as a chain of coupled cells where the wall between cells has been completely removed to increase the coupling (Fig. 10), leaving around the axis only a series of hollow tubes called 'drift tubes'. A high coupling offers the advantage of a large bandwidth, with sufficient spacing between the modes to avoid dangerous instabilities even when the chain is made of a large number of cells. Moreover, in the particular case of a structure operating in the 0-mode, removing the cell-to-cell walls does not influence power loss because the RF currents supporting this specific mode flow only on the external tank and on the tubes. The drift tubes have two functions. The first is to hide the particles during the half RF period when the electric field on the axis is decelerating, and the second is to concentrate the electric field around the gap to increase the acceleration efficiency. On top of that, if the diameter of the drift tubes is sufficiently large they can be used to house the focusing quadrupoles that at low energy are required to keep the beam transversally focused. The drift tubes are suspended from the outer tank by means of a supporting stem. The basic DTL structure is shown in Fig. 11.

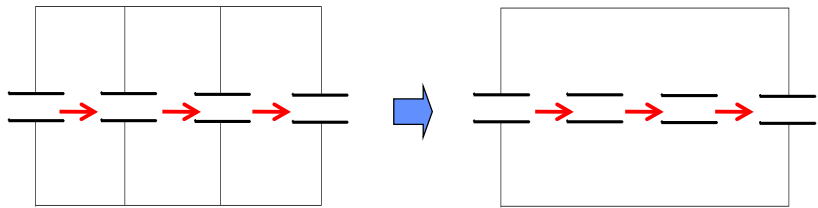

**Fig. 10**: A chain of resonators operating in 0-mode in a DTL. The arrows indicate the direction of the electric field on the axis.

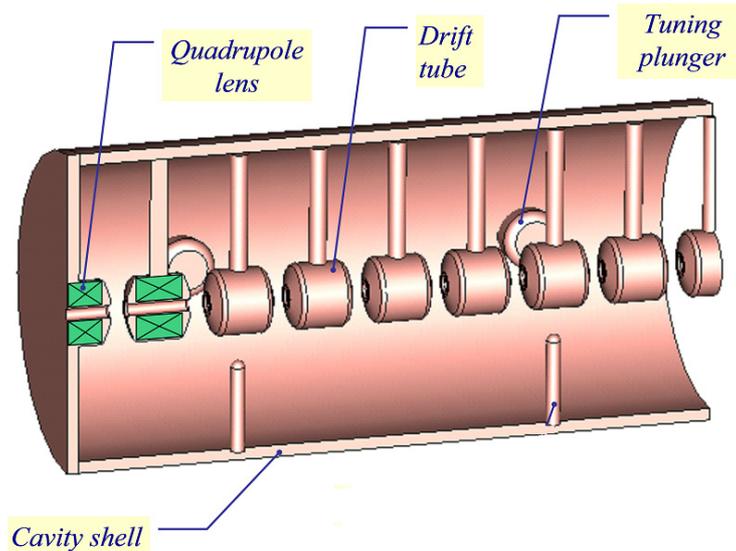

**Fig. 11**: DTL structure

An important feature of the DTL is that the length of the cells and of the drift tubes can be easily increased along the structure to follow the increase in beam velocity without influencing the frequency or the electric field distribution. In the theoretical approach developed in Section 2, all relations depend only on the frequency and not on the inductance or capacitance of the single cells. If the capacitance and inductance from one cell to another is changed, keeping constant their product and therefore the cell frequency, all of the relations developed in Section 2 remain valid; the mode frequencies and the relative amplitudes in the cells will not change. Therefore, if the length of the cells is progressively increased keeping constant their frequency, the system will still behave in the standard way and the operating 0-mode will keep all of its properties. Tuning cells of different length to the same frequency is almost straightforward because increasing the cell and drift tube lengths by the same proportion, the inductance will increase (longer cells) and the capacitance will decrease (larger gaps) by the same amount; at a first approximation the variations will compensate, keeping the frequency constant. Only minor adjustments to the gap lengths will be required to compensate for second-order effects.

The possibility to adjust each individual cell length to the particle $\beta$ together with the option of easily inserting focusing quadrupoles in the structure makes the DTL an ideal structure for initial acceleration in a proton linac, for energies from a few MeV to some 50–100 MeV. As an example, Fig. 12 shows a three-dimensional open view of the CERN Linac4 DTL, which will accelerate H⁻ ions (protons with two electrons) from 3 MeV to 50 MeV. The structure is divided into three individual 352.2 MHz resonators, for a total of 120 cells in a length of 19 m. The relativistic velocity increases from $\beta = 0.08$ to $\beta = 0.31$, and correspondingly the cell length $\beta\lambda$ increases by a factor of 3.9.

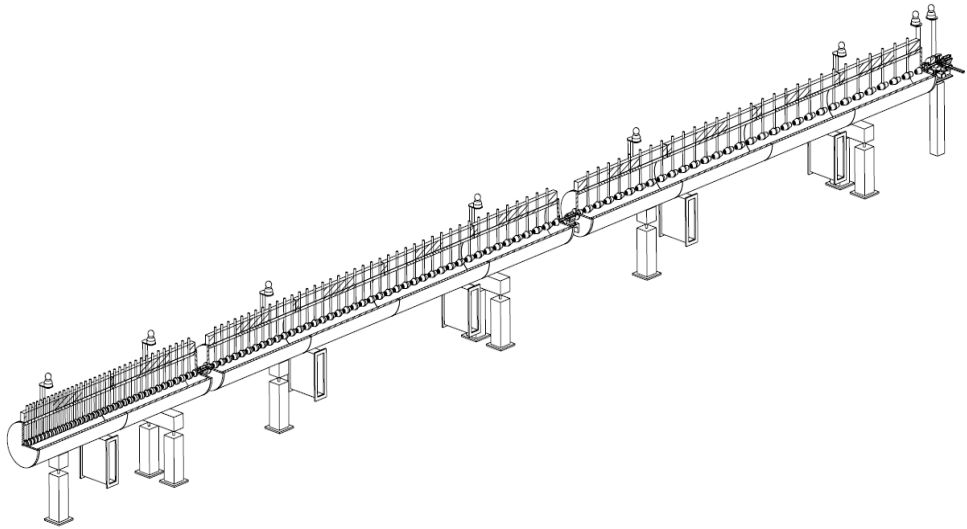

**Fig. 12**: Three-dimensional open view of the CERN Linac4 DTL

## 5   π-mode structures: the PI-Mode Structure (PIMS) and elliptical cavities

Structures operating in the π mode are widely used, in the magnetic coupled normally conducting version and in the electrically-coupled superconducting version. The coupling is provided by a slot on the external wall in the first case or by a sufficiently large opening on the axis in the second case. In both cases the cell length is kept constant inside short cavities made of a few (4 to 10, depending on the specific application) identical cells. Varying the cell length inside the cavities would complicate the design, because for the π-mode not just the cell frequency but also the coupling factor depends on the cell length, and would increase the construction cost. For these reasons, π-mode structures are commonly used in the high-energy range of a proton linear accelerator for energies above about 100 MeV, where the beam phase slippage is small.

As an example of a normally conducting π-mode structure, Fig. 13 shows the PI-Mode Structure (PIMS) that is being built at CERN for Linac4. Resonating at 352.2 MHz, it covers the energy range between 100 MeV and 160 MeV. The PIMS cavities are made of seven cells, coupled via two slots in the connecting wall (visible in Fig. 13(a)); the pairs of slots on the two sides of a cell are rotated by 90° to minimize second-neighbour couplings that could perturb the dispersion curve. The complete PIMS section is made of 12 seven-cell cavities. While the cell length inside each cavity is constant, it increases from cavity to cavity, matching the increase in $\beta$.

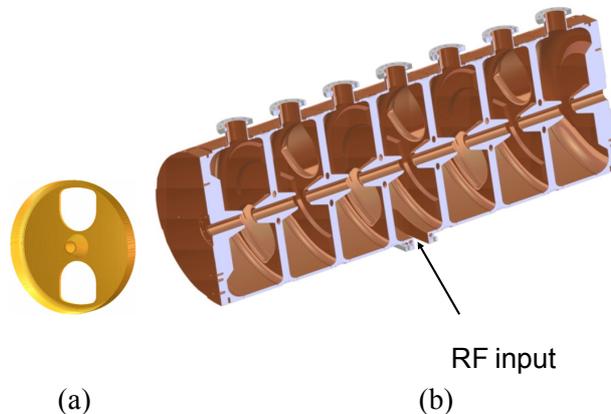

(a)                              (b)
**Fig. 13**: The PIMS seven-cell cavity. (a) Coupling slots; (b) cell structure

Figure 14 shows a typical superconducting low-$\beta$ cavity operating in the π mode. This particular cavity is made of five cells of identical length; the field distribution inside this type of cavities is shown in Fig. 5.

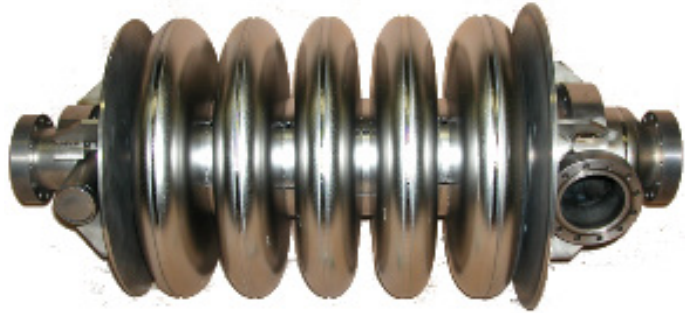

**Fig. 14**: A five-cell elliptical superconducting cavity

## 6   π/2-mode structures

When the chain of coupled resonators becomes very long, there is some interest in operating in the less efficient π/2-mode to reduce asymmetries in the electric field on the axis. This mode allows stable operation of long chains of coupled cells, which can then be fed by single high-power RF sources, which are less expensive than many smaller power units.

In Fig. 8 we see that the bandwidth of a coupled system (difference between maximum and minimum mode frequency) is proportional to the coupling factor and independent of the number of cells. Therefore, if we have a large number of cells and a large number of modes on the dispersion curve the modes will be very close in frequency to each other; this is particularly true for the 0 and π modes that lie in a region of the curve where the derivative is small. The modes will remain separated because their $Q$ value is usually sufficiently high; however, an important consequence of having several other modes close in frequency to the operating mode is that the system becomes extremely sensitive to mechanical errors. Small deviations from the design frequency in some cells in the chain (coming, for example, from machining errors) change the boundary conditions of the operating mode, forcing the system to introduce components from the adjacent modes to respect the new perturbed boundary conditions. These components are inversely proportional to the difference in frequency between operating and perturbing modes, making long structures more sensitive to errors than shorter ones.

In the π/2 mode instead, not only is the distance in frequency between the operating mode and the perturbing modes the largest, but their effect is compensated for, making the chain of resonators virtually insensitive to mechanical errors. The reason is that the components from perturbing modes add up to the field distribution of the operating mode with a positive or negative sign, depending upon whether they are higher or lower in frequency than the operating mode. Observing that the modes on the two sides of the operating π/2 mode have the same field distributions in the cells that are excited (but different ones in the cells that are empty), an error in the chain of resonators will be compensated for by symmetric components of the modes higher or lower in frequency. In principle, a π/2-mode structure can be totally insensitive to errors; in practice, this requires a perfect symmetry of the perturbing modes around the operating one, which is usually difficult to achieve. Reduction of the error sensitivity between a factor of 10 and 100 when going from a 0 or π mode to a π/2 mode is usually considered to be satisfactory.

The best known π/2 structure is the Side-Coupled Linac (SCL) structure (Fig. 15) developed at the Los Alamos National Laboratories in the 1960s. Here the coupling is magnetic, through slots on the cell walls, and the coupling cells are moved away from the beam axis and placed symmetrically on both sides of the chain of accelerating cells. The result is that, from the electromagnetic point of view, the structure operates in the π/2 mode providing stabilization of the field, whereas the beam travelling on the axis sees the typical field distribution of a π mode with maximum acceleration. Side-coupled structures are used at high energy and high RF frequency (from about 700 MHz), where high-power klystrons provide an economical way to feed a large number of cells for which operation in 0 or π mode would be impossible because of the strong sensitivity to mechanical errors. This particular type of structure is commonly used in the large number of commercially produced electron linacs for radiotherapy, making it the most widespread linac structure in the world.

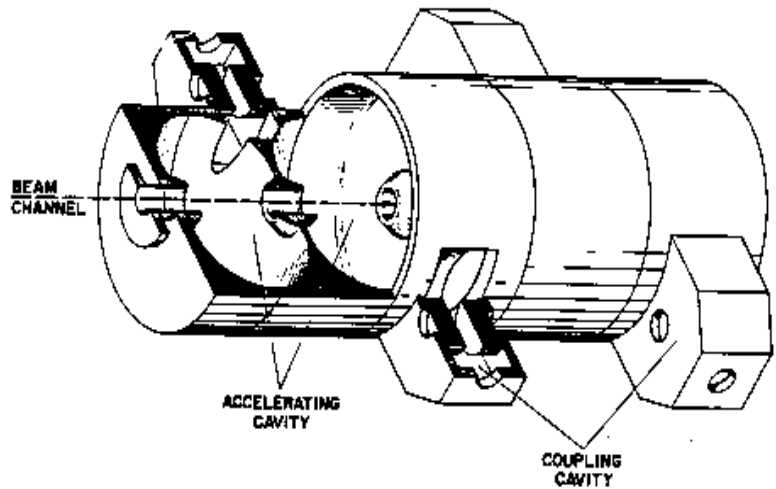

**Fig. 15**: The SCL structure [3]

## 7  Travelling wave structures, electron linacs

The treatment developed in the sections above concerns standing wave structures, where the RF power is injected at some position inside a closed multi-cell resonant structure and excites a given field pattern; after switching the RF power on, the fields increase inside the structure until reaching an equilibrium where all of the power injected is dissipated via ohmic losses in the metallic walls. Standing-wave operation is mandatory for proton and heavy ion linacs because it allows adaptation of the length of the cells to the changing particle velocity.

Electrons, on the other hand, become almost immediately relativistic (see Fig. 1) and can use for acceleration cells of identical length; on top of that, the beam focusing required for electrons is lower, making it possible to use much longer accelerating structures than that required for protons. Under these conditions, acceleration in the so-called travelling-wave mode is more efficient than in a standing-wave mode.

A structure operating in travelling-wave mode is again made of a sequence of cells, electrically or magnetically coupled. The most common structure used for electrons in travelling-wave mode is a sequence of cells electrically coupled on the beam axis, as in Fig. 16. This is a coupled-cell system of the type introduced in Section 3 and can be described by the same mathematical treatment; in order to obtain travelling-wave mode operation we have to consider a case where the chain of cells is 'infinite', i.e. with open boundary conditions at the two ends. In a real accelerator, one can simulate an infinite structure by injecting the RF power at one end of the structure and extracting it into a matched load at the other end (Fig. 17); in this way, the RF wave coming from the generator will travel inside the structure and will be absorbed by the load without any reflection, as would be the case for an infinitely long structure.

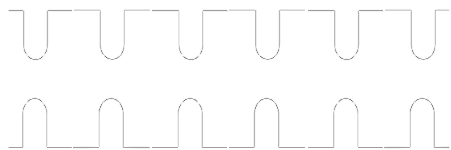

**Fig. 16**: Electrically-coupled cells

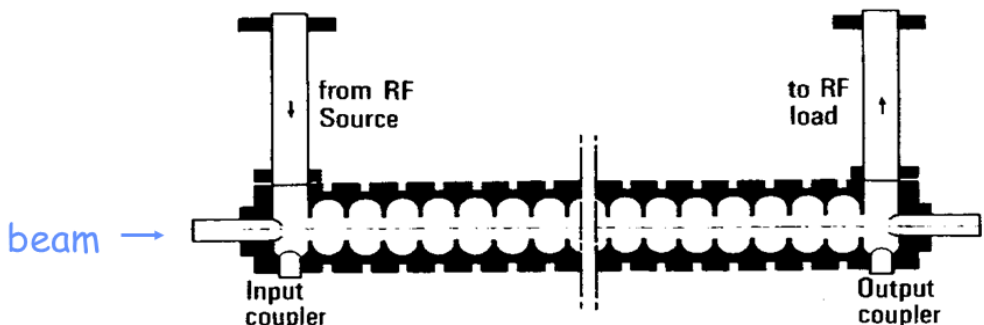

**Fig. 17**: Travelling-wave coupled-cell structure for electrons

In Eq. (12) for $N \to \infty$ the phase difference $\pi/N$ between modes becomes infinitely small and the number of modes in the dispersion curve in Fig. 8 becomes infinite. All the modes between 0 and $\pi$ are now permitted, each mode being still characterized by a phase difference between cells $\Phi_q$ varying between 0 and $\pi$. Finding the corresponding field amplitude is more complicated than in the standing wave case, the order of the matrix M going to infinity. The expression for the electric field, however, is similar to that of Eq. (14), considering that for cells of identical length $i = z/d$, with $z$ the longitudinal position along the structure. The solution for the travelling-wave case is

$$E(z,t) = E_0 \, e^{j\left(\omega t - \frac{\Phi}{d} z\right)} . \qquad (18)$$

This relation represents a wave travelling in the positive $z$ direction with velocity

$$v_\Phi = \frac{\omega}{\Phi} d . \qquad (19)$$

Here $v_\Phi$ is the phase velocity of the wave, the velocity at which a given phase is seen travelling in the $z$ direction. It is interesting at this point to plot the dispersion curve of the system presented in Fig. 8 as a function of $\Phi/d$, the 'wave number' of the travelling wave, to obtain the curve of Fig. 18.

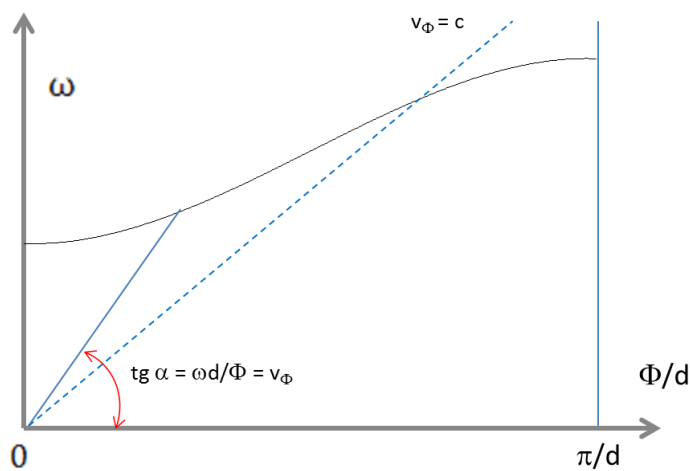

**Fig. 18**: Dispersion curve of a travelling-wave coupled-cell structure for electrons

Because a given phase of the sinusoidal electric field will travel with the phase velocity, if a particle enters the structure on the maximum electric field and then travels at the same velocity of the wave it will see the maximum electric field all along the structure and will receive maximum acceleration. It can be observed that in Fig. 18 the phase velocity for a given phase $\Phi$ is the tangent of the angle $\alpha$ between the horizontal axis and the point corresponding to $\Phi$ on the curve. The phase velocity is therefore infinite for $\Phi = 0$ and then decreases until a minimum reached for $\Phi = \pi$. This does not mean that there is something in the structure travelling at 'infinite' velocity, higher than the speed of light. The phase velocity is only the apparent velocity of a wave parameter; the information and the energy carried by the wave travel at the 'group velocity' that for a given $\Phi$ is equal to the tangent of the curve in Fig. 18. The group velocity starts at zero, increases up to a value always below the speed of light and then decreases again to zero.

Assuming that we want to accelerate a relativistic electron beam of velocity $c$, it is important to see if and where the phase velocity can be equal to $c$. The curve $v_\Phi = c$ is a straight line on the plot in Fig. 18 and there is a well-defined position where the $v_\Phi = c$ line intersects the dispersion curve. If we excite a wave at the corresponding frequency inside the structure, the wave will propagate at exactly the speed of light, remaining in phase with a bunch of relativistic particles, thus providing maximum acceleration. Common electron linacs usually adopt a frequency of 3 GHz; for a typical cell length $d = 10$ cm and the coupling factors achievable with on-axis electric coupling the $\Phi = 2\pi/3$ mode can be conveniently used for acceleration providing at the same time a sufficiently high group velocity.

Travelling wave modes present a high efficiency in the transfer of energy between the wave and the beam, but dissipate on the load at the end of the structure the fraction of the RF power that is not lost by ohmic loss on the walls. Their final accelerating efficiency is not much different from standing-wave structures for $\beta = 1$; the dimensions and the design details have more influence on the efficiency than the mode of operation.

## 8    Proton accelerating structures: comparison of shunt impedances

We have so far considered the three main groups of normally conducting coupled-cell accelerating structures for protons: the 0 mode, $\pi$ mode, and $\pi/2$ mode, and described the main representative of each category, the DTL, PIMS and SCL, respectively. Many more structures are used in linacs, which can be variants of these main families, mix properties of different families or even be based on different operating modes and different approaches. Among the best known 'alternative' structures are the Separated-DTL (SDTL), a variant of the DTL without quadrupoles in the drift tubes, the Annular Coupled Structure (ACS), a $\pi/2$-mode structure with a different coupling cell geometry from the SCL, and the Cell-Coupled DTL (CCDTL), a structure mixing a 0-mode with a $\pi/2$-mode operation, etc.

A particular category of linac structures is based on operation in a transverse electric (TE) mode instead of the usual transverse magnetic TM mode. The TE modes are called H modes in the German literature, and these structures are usually referred to as 'H-mode' structures. A TE mode in principle has an electric field only in the transverse direction and therefore cannot be used for acceleration. If drift tubes are placed in the structure connected alternatively to two sides of the resonator, however, the electric field of the TE11 mode can be forced in the longitudinal direction between the drift tubes and thus be able to provide acceleration. These are the so-called 'IH structures'. In a similar way, if the supports of the drift tubes are placed on the two transverse axes of the accelerating structure the TE21 mode can be forced to have a longitudinal electric field between the drift tubes: this is the 'CH structure'. IH and CH are very compact in terms of transverse dimensions; this is an advantage for low frequencies, but makes their construction difficult if high frequencies are required.

The choice of the most appropriate accelerating structure for a given project is very complex, being based on the comparison of many parameters. One of the most important figures of merit used for the selection of the accelerating structure is the shunt impedance, which represents the efficiency of an RF cavity in converting RF power into voltage across a gap. This is defined as

$$Z = \frac{V_0^2}{P} \qquad (20)$$

where $V_0$ is the peak RF voltage in a gap, and $P$ the RF power dissipated on the cavity walls to establish the voltage $V_0$. When the reference is to the effective voltage seen by a particle crossing the gap at velocity $\beta c$, we define the effective shunt impedance as

$$ZT^2 = \frac{(V_0 T)^2}{P} \qquad (21)$$

with $T$ the transit time factor of the particle crossing the gap (ratio of voltage seen by the particle crossing the gap over maximum voltage available). If the structure has many gaps, we can refer to the shunt impedance per unit length, usually expressed in ohms per metre. It must be noted that here we use the 'linac' definition, considering the shunt impedance as a sort of efficiency, i.e. a ratio between useful work (the voltage available to the beam, which is proportional to the energy gained by a particle, squared for dimensional reasons) and the energy (in this case power) required to obtain it. If instead we start from the consideration that the shunt impedance is the equivalent resistance in the parallel equivalent circuit of a cavity resonator, we need to add a factor of 2 at the denominator of the previous relations. This is the 'circuit' or 'RF' definition of shunt impedance.

RF power is expensive, and the goal of every designer of normally conducting accelerating structures is to maximize the shunt impedance; this is not straightforward, because this parameter depends on the mode used for acceleration, on the frequency and on the geometry of the structure. Other considerations of course come into play in the overall optimization; however, the shunt impedance remains one of the essential references for the structure designer.

Comparing structures in terms of shunt impedance is not easy, because frequency and geometry are strictly related to a specific project; however, a study made in 2008 by the EU-funded Joint Research Activity High-Intensity Pulsed Power Injectors (HIPPI) attempted at making a comparison as objective as possible for several types of structures, coming to the shunt impedance curves presented in Fig. 19. Here are compared eight different designs being studied in three different European Laboratories [4].

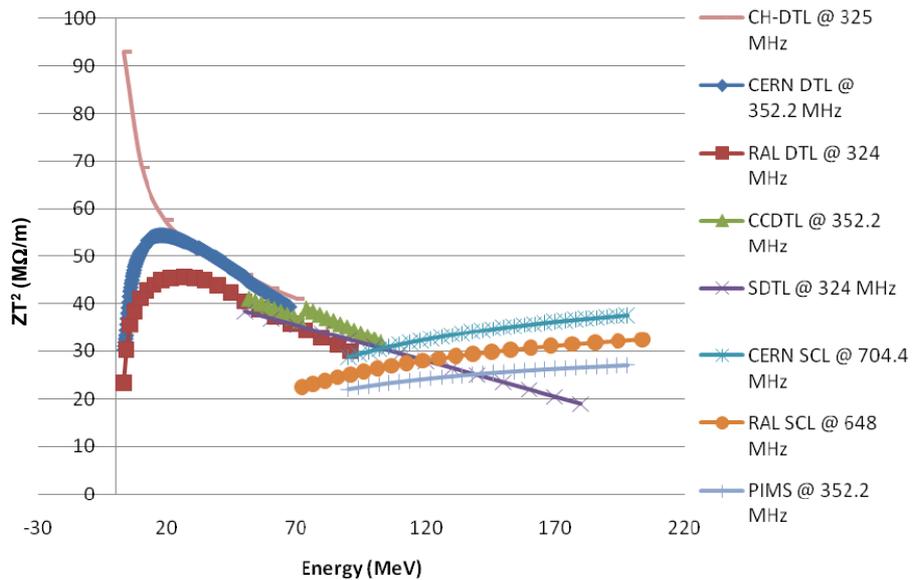

**Fig. 19**: Shunt impedance curves for different low-$\beta$ structures

The values presented here correspond to simulations of effective shunt impedance per unit length corrected for the additional losses expected in the real case and to optimised designs. The structures taken into consideration belong to two frequency ranges: the 324–352 MHz range and its double harmonic, 648–704 MHz. Higher operating frequencies have inherently higher shunt impedance; however, beam dynamics requirements in the first low-energy stages impose starting acceleration at frequencies below about 400 MHz.

From the curves it appears that for all structures the shunt impedance has a more or less pronounced dependence on beam energy, due to the different distribution of RF currents and losses in cells of different length corresponding to different $\beta$. Whereas 0-mode structures (DTL, but also the CCDTL in this context) have maximum shunt impedance around 20–30 MeV and then show a rapid decrease of the efficiency with energy, the π-mode structures' shunt impedance increases with energy, but from lower values than 0-mode structures. A natural transition point between these two types of structures would be around 100 MeV. For π-mode structures, remaining at the basic RF frequency leads to about 25% lower shunt impedance than doubling the frequency (comparing CERN SCL and PIMS curves). Different considerations apply to H-mode structures; the CH considered in this comparison has by far the highest shunt impedance below 20 MeV while above its behaviour is similar to that of TM 0-mode structures.

## 9  Low-$\beta$ superconducting structures

For superconducting structures, shunt impedance and power dissipation are not a concern, and the much lower RF power allows the use of simpler and relatively inexpensive amplifiers. A separated-cavity configuration such as Fig. 3 is therefore preferred for most superconducting linac applications at low energy, up to some 100–150 MeV, where more operational flexibility is required and where the short cavity lengths allow the ability to have more quadrupoles per unit length, as required at low energy. At higher energies, superconducting linacs use multi-cell π-mode cavities those presented in Fig. 12.

We must, however, observe that only few low-$\beta$ linacs use single-gap cavities; even for superconducting structures, economic reasons suggest adopting structures with generally two or, in some cases, three or four gaps, with an inherent phase slippage that remains, however, small and acceptable for a reduced number of gaps. In particular, the most widespread resonator used for very low-beta heavy ion applications is the quarter-wavelength resonator (QWR), sometimes declined in the half-wave resonator (HWR) form, when it is important to avoid even small dipole field components on the axis.

A particular structure recently developed for high duty-cycle proton beam applications at medium energy is the spoke resonator. In this cavity the electric field across the gaps is generated by a magnetic field turning around the supports, which are known as spokes. Its main advantages are the compact dimensions and the relative insensitivity to mechanical vibrations. Similarly, for intense proton or deuteron beams is proposed a superconducting version of the CH resonator. Some examples of these structures are presented in Fig. 20.

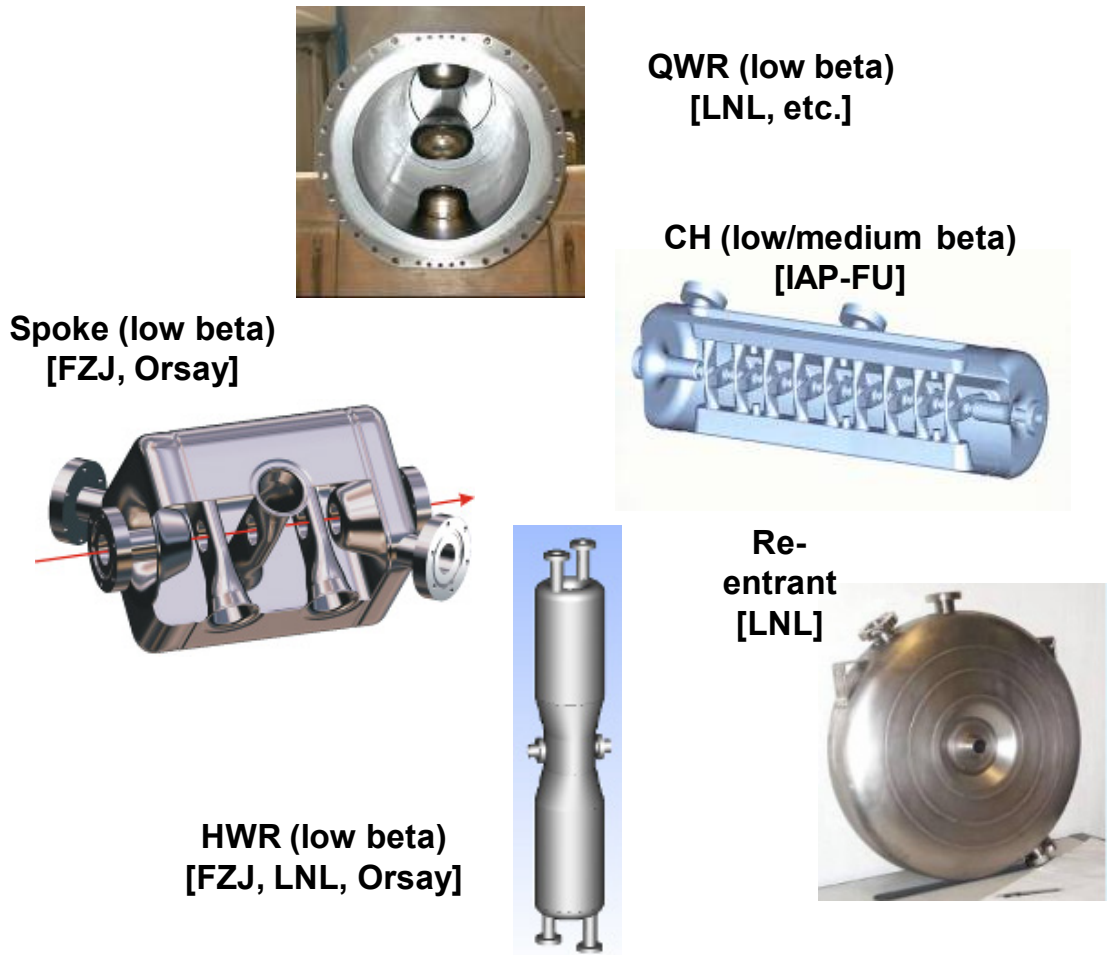

**Fig. 20**: Some examples of superconducting low-$\beta$ structures

## 10 Beam dynamics in linear accelerators

### 10.1 Longitudinal plane

We have seen that bunches of particles must be synchronous with the accelerating wave to achieve the maximum acceleration. This means that they have to be injected into the linac on a well-defined phase with respect to the accelerating sinusoidal field, and then they need to maintain this phase during the acceleration process. Linac beams are usually made of a large number of particles with a given spread in phase and in energy. If the injection phase corresponds to the crest of the wave ($\varphi = 0º$ in the linac definition) for maximum acceleration, particles having slightly higher or lower phases will gain less energy. They will slowly lose synchronicity until they are lost.

In linacs, the same principle of phase stability holds as in synchrotrons: if the injected beam is not centred on the crest of the wave but around a slightly lower phase, a 'synchronous phase' $\varphi_s$ whose typical values are between −20º and −30º, particles that are not on the central phase will oscillate around the synchronous phase during the acceleration process. The resulting longitudinal motion is confined, and the oscillation is represented by an elliptical motion of each particle in the longitudinal phase plane, i.e. the plane ($\Delta\varphi, \Delta W$) of phase and energy difference with respect to the synchronous particle. The relation between the synchronous phase in an accelerating sinusoidal field, and the longitudinal phase plane is presented in Fig. 21.

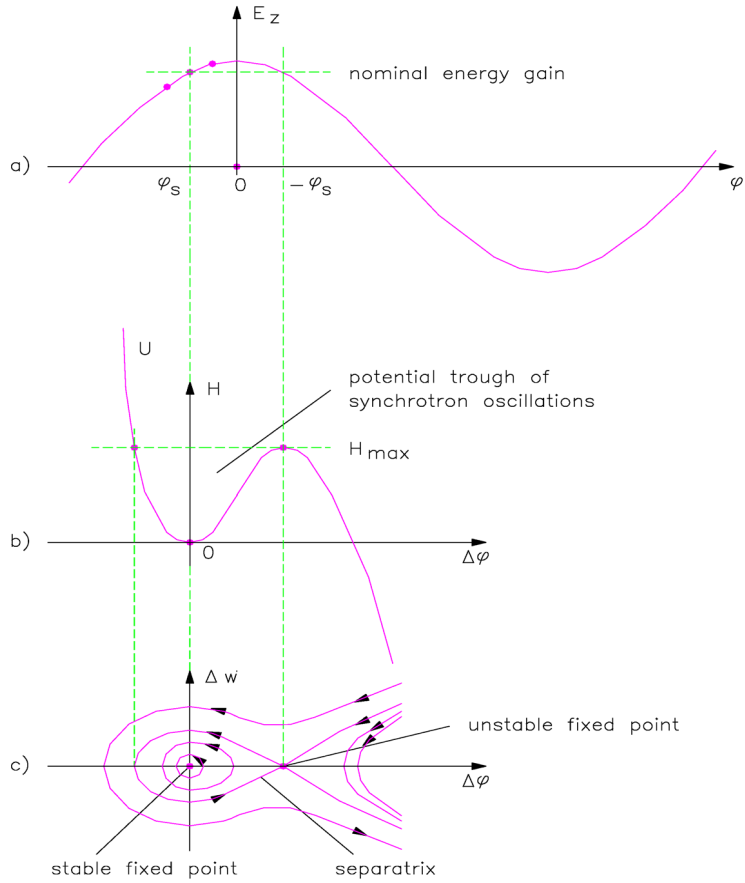

**Fig. 21:** Longitudinal motion of an ion beam

It is interesting to observe that the frequency of longitudinal oscillations, i.e. the number of oscillations in the longitudinal phase plane per unit time, depends on the velocity of the beam. A simple approximate formula for the frequency of small oscillations $\omega_l$ can be found, for example, in Ref. [5],

$$\omega_l^2 = \omega_0^2 \frac{qE_0 T \sin(-\phi)\lambda}{2\pi mc^2 \beta\gamma^3} \quad . \tag{22}$$

Here $\omega_0$ and $\lambda$ are the RF frequency and wavelength, respectively, $E_0 T$ is the effective accelerating gradient and $\varphi$ is the synchronous phase. The oscillation frequency is proportional to $1/\beta\gamma^3$: when the beam becomes relativistic, the oscillation frequency decreases rapidly. At the limit of $\beta\gamma^3 \gg 1$ the oscillations will stop and the beam is practically 'frozen' in phase and in energy with respect to the synchronous particle. For example, in a proton linac $1/\beta\gamma^3$ and correspondingly $\omega_l$ can decrease by two or three orders of magnitude from the beginning of the acceleration to the high-energy section.

Another important relativistic effect for ion beams is the 'phase damping', the shortening of bunch length in the longitudinal plane. This can be understood considering that, as the beam becomes more relativistic, its length in $z$ seen by an external observer will contract due to relativity. A precise relativistic calculation shows that the phase damping is proportional to $1/(\beta\gamma)^{3/4}$

$$\Delta\phi = \frac{\text{const}}{(\beta\gamma)^{3/4}} \quad . \tag{23}$$

When a beam becomes relativistic, not only will its longitudinal oscillations slow down, but the bunch will also compact around the centre particle.

## 10.2 Transverse plane

Transversally in a linac, the beam will be subject to an external focusing force, provided by an array of quadrupoles or solenoids. This force has to counteract the defocusing forces that either develop inside the particle beam or come from the interaction with the accelerating field. The main defocusing contributions come from space charge forces and from RF defocusing.

### *10.2.1 Space charge forces*

Space charge forces represent the Coulomb repulsion inside the bunch between particles of the same sign. In the case of high-intensity linacs at low energy, space charge forces are one of the main design concerns. At relativistic velocity, however, the space charge repulsion starts to be compensated for by the attraction due to the magnetic field generated by the beam, and finally disappears at the limit $v = c$. Space charge forces can be calculated only for very simple cases, such as that of an infinitely long cylindrical bunch with density $n(r)$ travelling at velocity $v$ (Fig. 22).

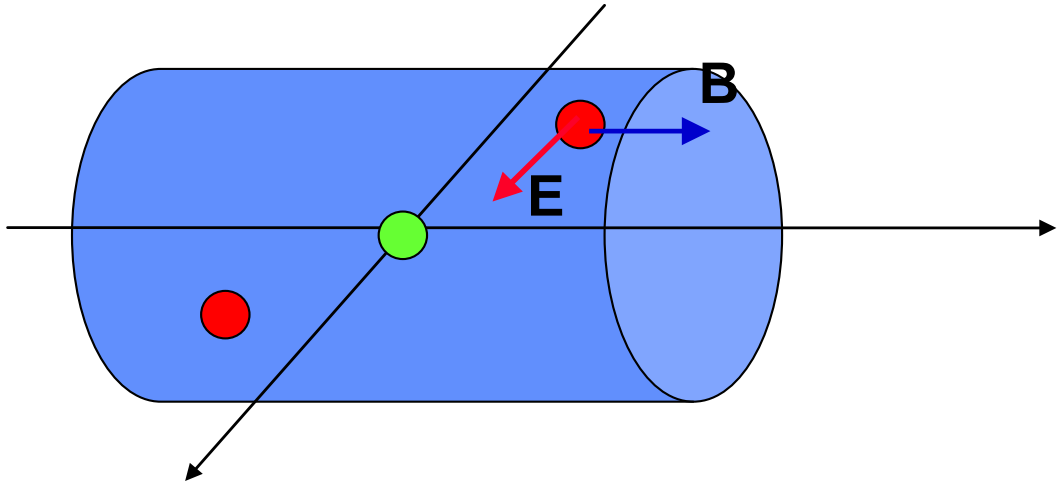

**Fig. 22:** Forces acting on a particle inside an infinitely long bunch

In this case, the electric and magnetic fields active on a particle at distance $r$ from the axis can be written as

$$E_r = \frac{e}{2\pi\varepsilon\, r} \int_0^r n(r)\, r\, dr \quad . \tag{24}$$

The resulting overall force on a particle in the bunch is orientated in the radial direction, and has intensity

$$F = e(E_r - vB_\phi) = eE_r\left(1 - \frac{v^2}{c^2}\right) = eE_r(1-\beta^2) = \frac{eE_r}{\gamma^2} \quad . \tag{25}$$

The overall space charge force is then proportional to $1/\gamma^2$ and will disappear for $\gamma \to \infty$.

### *10.2.2 RF defocusing forces*

RF defocusing is the transverse defocusing experienced by a particle that crosses an accelerating gap on a longitudinally focusing RF phase. We have seen in Section 10.1 that for longitudinal stability the beam will cross the gap when the field is increasing ($\varphi_s < 0$). Figure 23 shows a schematic configuration of the electric field in an accelerating gap. In correspondence to the entry and exit openings, the electric field has a transverse component, focusing at the entrance to the gap and defocusing at the exit, proportional to the distance from the axis. Because the field is increasing when the beam crosses the gap, the defocusing effect will be stronger than the focusing effect, and the net result will be a defocusing force proportional to the time required for the beam to cross the gap. A Lorentz transformation from the laboratory frame to the frame of the particles of the electric and magnetic field forces acting on a particle allows calculation of the radial momentum impulse per period. Carrying out this calculation, one can find

$$\Delta p_r = -\frac{\pi e E_0 T L r \sin\phi}{c \beta^2 \gamma^2 \lambda} \ . \tag{26}$$

Again, this effect is proportional to $1/\gamma^2$, and will disappear at high beam velocities.

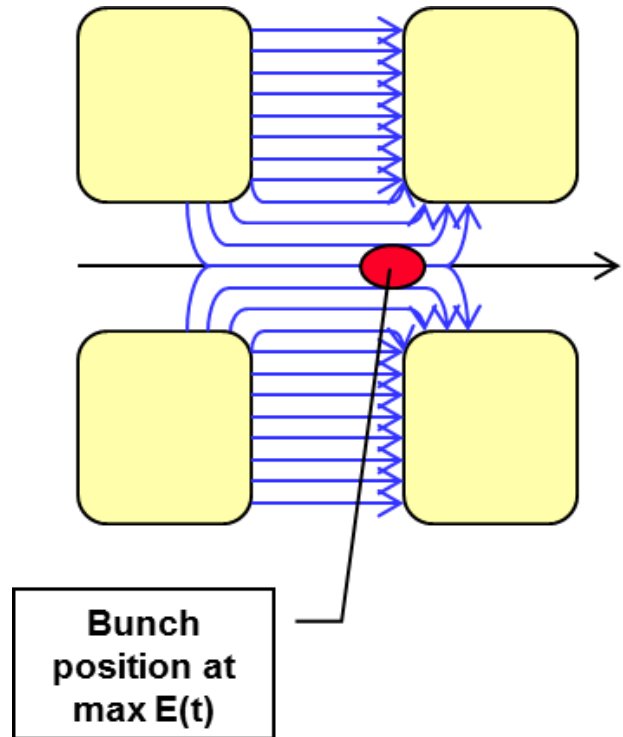

**Fig. 23:** Electric field line configuration around a gap and position of the bunch at maximum field

### 10.3 Transverse equilibrium

We have seen that at low energies in particular, strong transverse defocusing forces act on the beam; to transport it with minimum particle loss we need to compensate for the defocusing forces with external focusing forces, obtained by means of a standard alternating gradient focusing channel made of quadrupoles placed along the beam line. Because the linac is made of a series of accelerating structures the standard focusing solution consists (Fig. 24) of alternating the accelerating sections with focusing sections made of one quadrupole (singlet focusing), two quadrupoles (doublet focusing) or three quadrupoles (triplet focusing). This layout defines a *focusing period*, corresponding to the length after which the structure repeats itself. It is important to observe that because the accelerating sections have to match the increasing beam velocity, the accelerating structures can have increasing lengths and therefore the basic focusing period does not necessarily have a constant geometrical length; nevertheless, the travel time of the beam within a focusing period remains constant. The maximum distance between the focusing elements depends on the beam energy, as we will see in the following; it goes from only one gap in the DTL to one or more structures containing many gaps at high energies.

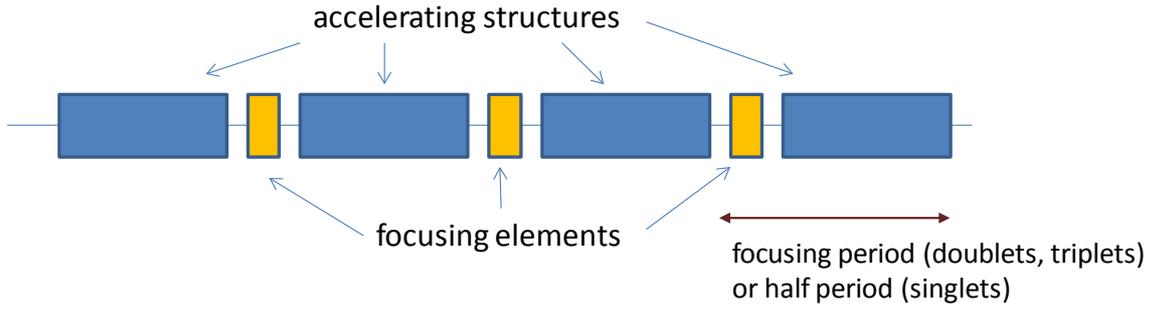

**Fig. 24**: Standard linac section

The beam must be transversally in equilibrium between the external focusing forces and the internal defocusing forces; the equilibrium will necessarily be dynamic, resulting in an oscillation in time and in space of the beam parameters with a frequency that depends on the ratio between focusing and defocusing forces. The oscillation can be, as usual, decomposed into two independent oscillations with the same frequency in the transverse planes $x$ and $y$, the reference oscillating parameter being, as usual, the maximum beam radius. The oscillation is characterized by its frequency; instead of defining it with respect to time it is convenient to define the oscillation frequency in terms of the phase advance $\sigma_t$, the increase in the phase of the oscillation over a focusing period of the structure, usually constant or changing only smoothly in a linac section. Alternatively, the phase advance per unit length $k_t$ can be used; if $L$ is the period length of the focusing structure, then $k_t = \sigma_t/L$. Modern beam dynamics simulation code allow us to easily calculate the phase advance for a given focusing period, accelerating structure and quadrupole gradient; it is important to select accurately this parameter to minimize beam loss and to avoid excessive transverse emittance growth. The first basic rule is that $\sigma_t$ should always be <90º, to avoid resonances that could lead to emittance growth and to beam loss; it should also be higher than around 20º, to avoid the amplitude of the oscillations becoming too high and the beam size becoming too large. It should be emphasized that beam aperture is expensive; the effective shunt impedance introduced in Section 8 strongly decreases when increasing the aperture in the accelerating structures.

We can find an approximate relationship for the phase advance as a function of focusing and defocusing forces when referring to a simple theoretical case. First of all, one has to limit the analysis to beam oscillations in a simple F0D0 quadrupole lattice (focusing-drift–defocusing-drift, corresponding to the 'singlet' focusing case of Fig. 24) under smooth focusing approximation, i.e. averaging the localized effect of the focusing elements. Then, adding together the focusing and RF defocusing contributions to phase advance as derived, for example, in Ref. [5: Eq. (7.103)] and subtracting the space charge term as approximately calculated in the case of a uniform three-dimensional ellipsoidal bunch [5: Eq. (9.51)] we obtain for the phase advance per unit length

$$k_t^2 = \left(\frac{\sigma_t}{N\beta\lambda}\right)^2 = \left(\frac{qGl}{2mc\beta\gamma}\right)^2 - \frac{\pi q E_0 T \sin(-\phi)}{mc^2 \lambda \beta^3 \gamma^3} - \frac{3qI\lambda(1-f)}{8\pi\varepsilon_0 r_0^3 mc^3 \beta^2 \gamma^3}. \quad (27)$$

Here $N\beta\lambda$ is the length of the focusing period in units of $\beta\lambda$. The first term on the right-hand side of the equation is the focusing component: $Gl$ is the quadrupole integrated gradient, expressed as the product of gradient $G$ and length $l$ of the quadrupole. The second term is the RF defocusing: $E_0T$ is the effective accelerating gradient, $\lambda$ the RF wavelength, and $\varphi$ the synchronous phase. For $\varphi < 0$, corresponding to longitudinal stability, $\sin(-\varphi)$ is positive and this term is negative, i.e. defocusing. The third term is the approximate space charge contribution: $I$ is the beam current, $f$ is an ellipsoid form factor ($0 < f < 1$) and $r_0$ is the average beam radius. The other parameters in the equation define the particle and medium properties (charge $q$, mass $m$, relativistic parameters $\beta$ and $\gamma$, and free space permittivity $\varepsilon_0$).

This simple equation shows, although in an approximate simplified case, how the beam evolution in a linear accelerator depends on the delicate equilibrium between external focusing and internal defocusing forces. Real cases can only be solved numerically; however, the parametric dependence given by this equation remains valid, and allows us to determine how the beam dynamics will change with the particle $\beta$. At low velocities ($\beta \ll 1$, $\gamma \sim 1$) the defocusing terms are dominant. To keep the beam focused with a large enough phase advance per unit length one has to increase the integrated gradient $Gl$ and/or decrease the length of the focusing period $N\beta\lambda$, i.e. minimize the distance between focusing elements. This is, for example, the case of the Radio Frequency Quadrupole (RFQ), the structure of choice for low-energy ion beams (from $\beta \approx 0.01$ to $\beta \approx 0.1$). The RFQ provides a high focusing gradient by means of an electrostatic quadrupole field, with short cells at focusing period $\beta\lambda$. At higher energies, standard electromagnetic quadrupoles have a sufficiently high gradient, and a structure alternating accelerating gaps and quadrupoles can be used: the typical structure in this energy range is the DTL (Fig. 10), which has $2\beta\lambda$ focusing periods when focusing and defocusing quadrupoles alternate inside the drift tubes (F0D0). The gradient achievable in the DTL quadrupoles thanks to the large diameter of the drift tubes, together with the short focusing period, allows keeping the phase advance in an acceptable range even for high current and high space charge beams; for example, the CERN Linac2 can accelerate a beam current up to 180 mA, the maximum achieved so far in a DTL. As an example of DTL beam dynamics design, Fig. 25 presents quadrupole gradients and the corresponding phase advance for the CERN Linac4 DTL [6]. The corresponding oscillations of the beam envelope for the given input matching conditions are shown in Fig. 26 (the plot is for the maximum beam radius in $x$). As the phase advance decreases, the period of the oscillations becomes longer.

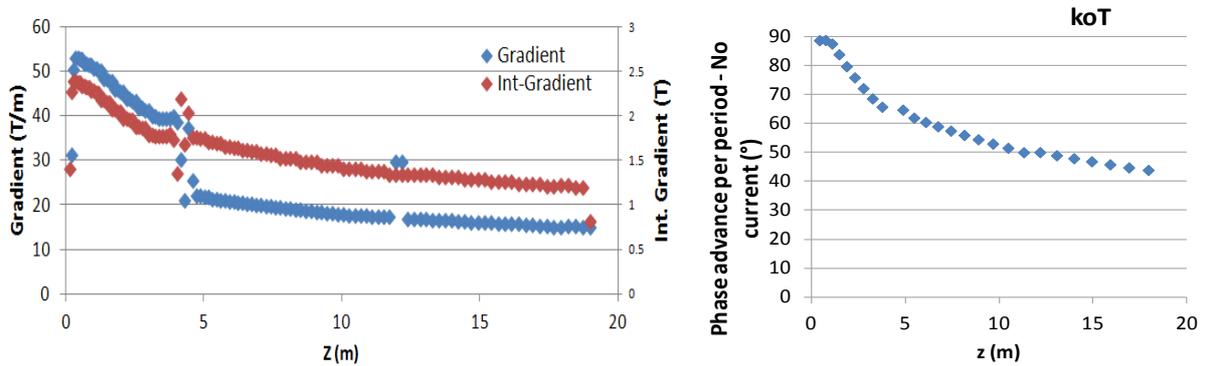

**Fig. 25:** Quadrupole gradients and corresponding phase advance for the Linac4 DTL design

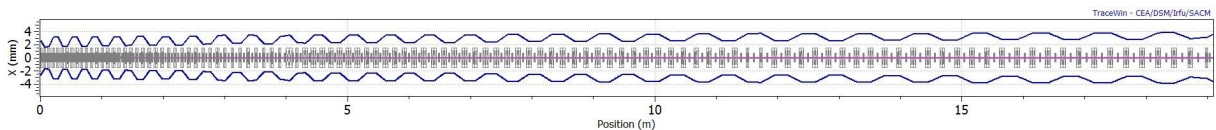

**Fig. 26:** Transverse root mean square beam envelope ($x$-plane) for the Linac4 DTL design

Going further up in energy, the defocusing terms (proportional to $1/\beta^3\gamma^3$ and $1/\beta^2\gamma^3$, respectively) decrease much faster than the focusing term (proportional to $1/\beta\gamma$). The focusing period can be increased, reducing the number of quadrupoles and simplifying the construction of the linac. Starting from energies between 50 MeV and 100 MeV, modern proton linacs adopt multi-cell structures operating in $\pi$ mode spaced by focusing quadrupoles; the structures can be normally conducting (Fig. 13) or superconducting (Fig. 14). In the CERN Linac4 design (45 keV to 160 MeV beam energy) the focusing period increases from $\beta\lambda$ in the RFQ to $15\beta\lambda$ in the last $\pi$-mode accelerating structure. The corresponding beam envelope is shown in Fig. 27. Heavy ions differ from protons in that usually ion currents are small and the space charge term can be neglected; immediately after the RFQ, the focusing period can go up to some units of $\beta\lambda$.

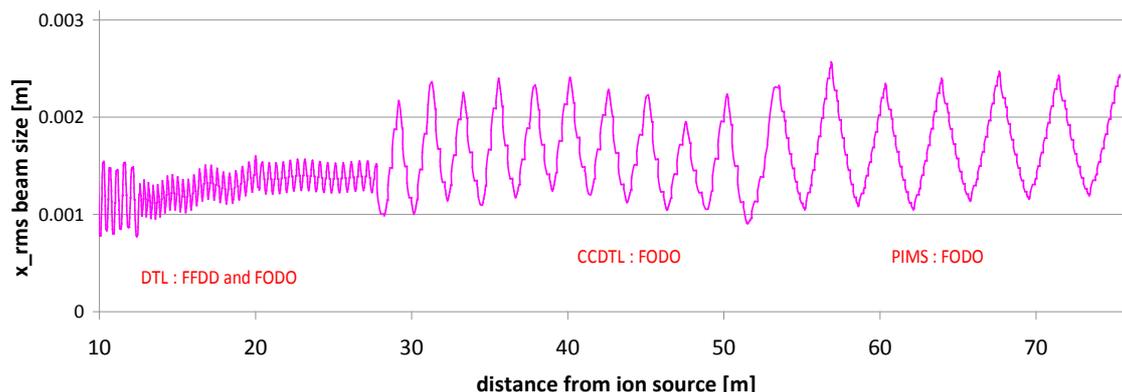

**Fig. 27:** Transverse root mean square beam envelope (*x*-plane) along the different Linac4 sections (3 MeV to 160 MeV).

Electrons present a simpler beam dynamic than protons or ions. Space charge and RF defocusing are present only in the very first stage of acceleration, at the exit of the electron gun and through the bunching section, where a beam dynamics approach similar to that for protons is used. As soon as acceleration starts the defocusing terms disappear; a quadrupole focusing system is still required to define phase advance and to compensate for instabilities but the distance between quadrupoles can be much larger than in the case of protons.

## 11  Linac architecture

From the previous sections, we can derive two basic rules defining the architecture of any proton or ion linear accelerator required to reach energies above a few MeV:

i) at low energy the linac cells have to be different in length to follow precisely the increase in beam velocity; at higher energy, the accelerating structures can be made out of sequences of identical cells, thus reducing the construction costs thanks to greater standardization and the use of longer vacuum structures;

ii) as the beam energy increases then less focusing is required because of the reduction in space charge and RF defocusing; at higher energy the focusing length can increase, reducing the number and cost of the quadrupoles.

These two rules lead to the same consequence: to keep construction and operation costs low a linac must change the type of structure and focusing scheme with the increase in energy, going from expensive structures integrating a large number of focusing elements that cover the low-energy range, up to more economical structures and focusing layouts covering the high-energy range. A linac will thus be made of different sections for different ranges of energy. After the ion source, the first accelerating structure will be a RFQ (described below); this is the only linac structure that can provide the strong focusing forces needed to compensate for space charge at low energy. The RFQ is an expensive structure that is not particularly efficient in using RF power; for this reason, from an energy of a few MeV it becomes convenient to go from the RFQ to another type of structure. Most linacs use a DTL for the following section; the DTL integrates in a single RF cavity cells of increasing length, matching precisely the increase in beam velocity, together with quadrupoles placed in every cell that provide a strong and uniform focusing. As an alternative to the DTL, linacs required to operate at high duty cycle (close to or at CW) and/or to provide acceleration of different ion types have an economic interest in using short superconducting structures, e.g. with two gaps at a fixed distance. Both a normally conducting DTL and a sequence of short superconducting cavities are quite expensive, and from energies between 50–100 MeV it is convenient to go to structures with many identical cells interleaved with quadrupoles. While at intermediate energies (50–200 MeV) special structures are still required, at higher energies long standardized modules of identical cells, often superconducting, can be used. Focusing is provided by quadrupoles placed between the modules.

As an example, Fig. 28 shows the layout of the CERN Linac4. This linac will cover the energy range between 45 keV (extraction from the ion source) and 160 MeV with four different types of accelerating structures, characterized by an increasing number of identical cells per accelerating structure and an increasing length of the focusing period (Table 1).

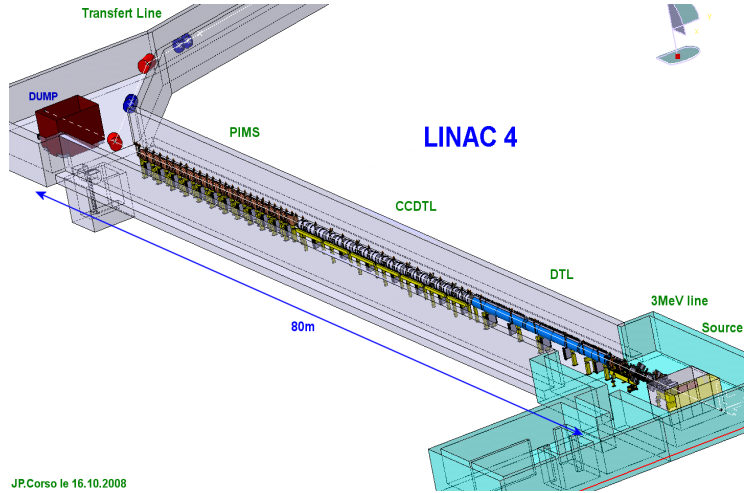

**Fig. 28**: Layout of Linac4 under construction at CERN

**Table 1:** Linac4 accelerating structure characteristics

| Section | Output energy (MeV) | Number of consecutive identical cells | Distance between quadropoles |
|---------|---------------------|---------------------------------------|------------------------------|
| RFQ     | 3                   | 1                                     | –                            |
| DTL     | 50                  | 1                                     | $\beta\lambda$               |
| CCDTL   | 100                 | 3                                     | $4\,\beta\lambda$            |
| PIMS    | 160                 | 7                                     | $5\,\beta\lambda$            |

## 11.1 The choice of the frequency

The choice of the appropriate frequency (or sequence of frequencies) for a linac has to take into account several factors coming from mechanical, RF, and beam dynamics considerations related to the different scaling with frequency of the linac parameters. First of all, accelerator dimensions and cell length are proportional to the wavelength $\lambda$: higher frequencies will result in smaller accelerating structures, requiring less copper or steel. Machining tolerances, however, scale as well as $\lambda$: smaller structures require tighter tolerances that are more expensive to achieve. From the RF point of view, higher frequencies are also preferable, because both the shunt impedance and the maximum surface electric field scale approximately as $\sqrt{f}$. (It should be noted that while it is evident that the peak surface electric field increases with frequency, its exact dependence and the limit for very high frequencies are still debated. While a dependence on the square root of the frequency is assumed in early studies, recent analysis indicates a linear dependence; a limitation related to surface damage appears for frequencies above about 10 GHz. The shunt impedance instead scales as the square root of the frequency in the case of a symmetric scaling of cavity dimensions; if the beam aperture is kept constant, as is usually the case, the dependence will be smaller.) In Section 8, however, we have seen that the RF defocusing scales with $1/\lambda$, becoming excessively high at high frequencies; RF defocusing and cell length in the RFQ usually define the maximum frequency that can be used in the initial section of a linac. A summary of the dependence with the frequency for the different linac parameters is given in Table 2.

**Table 2:** Scaling with frequency of some basic linear accelerator parameters

| Parameter | Scaling |
|---|---|
| RF defocusing | $\sim f$ |
| Cell length | $(\sim \beta\lambda) \sim 1/f$ |
| Peak electric field | $\sim \sqrt{f}$ |
| RF power efficiency (shunt impedance) | $\sim \sqrt{f}$ |
| Accelerator structure dimensions | $\sim 1/f$ |
| Machining tolerances | $\sim 1/f$ |

A first analysis of the frequency dependence indicates that high frequencies are economically convenient: the linac is shorter, it makes use of less RF power and can reach a higher accelerating field. Limitations to the frequency come from the mechanical construction costs that depend critically on the required tolerances and on the RF defocusing in the RFQ; for these reasons, modern linac designs tend to start with a basic frequency in the RFQ and then double it as soon as the cells become longer and the RF defocusing decreases. Doubling the frequency allows a clean bunch-to-bucket transfer, only half of the buckets being populated in the higher frequency structure, but requires some care in the longitudinal beam matching. The availability and cost of the RF power sources is an essential element in the choice of the frequency, and has to be considered carefully in particular when multiple frequencies are used.

All of these requirements result in some standardization in the frequencies commonly used in linacs: while in the past proton RFQs used to have frequencies around 200 MHz, nowadays frequencies in the range 325–425 MHz can be easily achieved. In the following linac sections, normally conducting or superconducting, a frequency jump of a factor of 2 is often applied, reaching the range 700–800 MHz.

## 11.2 Superconductivity and the warm–cold transition

A constant in the architecture of modern high-energy linacs is the use of superconductivity at high energy. The advantages of superconducting accelerating structures are evident: a much smaller RF system needs to deliver only the power directly going to the beam, a large beam aperture allows for a lower beam loss (although the particles in the beam halo are transported in the superconducting section and then lost in the following beam transport line) and, finally, the operating electricity costs are lower than for a normally conducting linac, a feature particularly important given the present concerns for energy saving. The energy argument is particularly important for high duty-cycle machines, where high power efficiency is an important requirement, and becomes proportionally less important for low-duty machines where the beam power is a minor fraction of the power required by the machine.

In defining an optimum linac design, however, some peculiar characteristics of superconducting systems have to be considered. A superconducting system needs a large cryogenic installation requiring a significant amount of power; for a low-duty linac this can be dominated by the static losses required for keeping the system at cryogenic temperature, leading to low overall power efficiency. Additionally, the requirement at low energy to provide many cold–warm transitions to accommodate the large number of warm quadrupoles required because of the short focusing periods increases the cost and complexity of the installation. On top of that, the difficulty in predicting the individual gradients of superconducting cavities makes them less attractive at low energy where, as we have seen, the sequence of cell lengths has to precisely follow the calculated increase in beam velocity.

The result is that while superconductivity is certainly the most attractive technology for linacs at high energy and high duty cycle, at low energy and low duty cycle the normally conducting structures remain more economical and more efficient. An exact comparison of cost and efficiency for the two technologies is difficult because it depends not only on energy and duty cycle, but also on other design parameters such as repetition frequency, peak beam current and pulse length. A high repetition frequency makes a superconducting linac less efficient, more power being lost during the long pulse rise time required to fill the superconducting cavities. The maximum current during pulse plays an important role, because whereas normally conducting linacs are more efficient operating with short pulses of high beam current, superconducting linacs prefer long pulses with less current, which requires a smaller RF installation. For all of these reasons, the optimum transition energy between warm and cold sections in a modern linac remains difficult to determine and requires a precise economical comparison of the two technologies for the parameters of each particular project. In CW linacs, however, the superconducting section can start immediately after the RFQ, usually at 3 MeV.

## 12   Low energy acceleration—the radio frequency quadrupole

The low-energy section, between the ion source and the first drift-tube-based accelerating structure, is probably the most challenging part of any hadron linear accelerator. It is in this part of the linac that:

i) Defocusing due to space charge forces is the highest. To compensate for space charge external focusing must be high, with short focusing periods and a large number of high gradient quadrupoles. The focusing achievable at low energy is however limited by the small dimensions of the accelerating cells: for example, in a DTL at 1 MeV, $\beta = 4.6\%$ and for $\lambda \sim 1$ m the maximum length of a quadrupole is one half of the period, i.e. about 20 mm, nearly the same as the required aperture. The quadrupole would be dominated by fringe fields and it would be impossible to achieve on the axis a gradient sufficient to control high space charge forces.

ii) The continuous beam coming out of the source has to be bunched in order to be accelerated in the first RF accelerating structure. The process of bunching by means of longitudinally focusing RF forces is a critical operation: it defines the longitudinal beam emittance and can lead to the loss of a large fraction of the particles if the resulting emittance is not matched to the acceptance of the first accelerating structure.

iii) Usual low-energy accelerating structures have reduced RF efficiency and high mechanical complexity, because of the need to adapt the length of every cell to the beam velocity. Short cells have high stray capacitances that for a given power dissipation reduce the effective voltage available for the beam, or in other terms they are particularly ineffective in concentrating on the axis the electric field requires for acceleration. The result is that the accelerator cost per metre (or per MeV acceleration) in this section tends to be the highest for the linac and needs to be carefully optimized.

Before the invention of the RFQ, the classical solution to cover this critical energy range was to extend as much as possible the extraction voltage from the ion source and to start the first accelerating structure, usually a DTL, from the lowest possible energy. The use of large HV generators, at the limit of technology, allowed the extraction from the source of a beam with sufficient velocity to be injected into a DTL of relatively low frequency (to increase $\lambda$) equipped with special short quadrupoles in the first drift tubes. In these systems, however, the maximum beam current was limited by the size and aperture of the first quadrupoles and by space charge in the transport line between the source and the DTL, and the low RF frequency reduced the overall acceleration efficiency. Bunching was provided by a single-gap RF cavity followed by a drift space before the DTL, a simple technique but presenting a low efficiency; only about 50% of the beam coming out of the ion source being injected in the longitudinal bucket at the first gap of the DTL.

These problems led to the development in the 1960s and 1970s of solutions to overcome the current limitations of conventional low-energy linac sections. A new revolutionary idea in this respect came from I. Kapchinsky of the Institute for Theoretical and Experimental Physics (ITEP) in Moscow, who proposed using at low energy an electric quadrupole focusing channel excited at RF frequency as an alternative to conventional electromagnetic quadrupoles. Electric quadrupole forces do not decrease at low particle velocity as does the Lorentz force of a magnetic quadrupole field; if the electric field is generated by an RF wave, a beam of particles travelling on the axis of the electric quadrupole will see an alternating gradient, resulting in a net focusing force. Kapchinsky's idea was to add to the quadrupole electrodes a longitudinal 'modulation' (i.e. a sinusoidal profile), which generates a longitudinal electric field component. By matching this longitudinal time-varying field with the velocity and phase of the particle beam it was possible to use this structure for bunching and for a moderate acceleration.

Starting from this idea, V. Teplyakov of the Institute for High Energy Physics (IHEP) in Protvino designed an RF resonator around the quadrupole electrodes, and the first paper on the Radio Frequency Quadrupole (RFQ) was published in Russia in 1969 [7]. Only a few years later the high-intensity proton linac team at Los Alamos Laboratories in the US noticed this paper and, adding some important improvements and building a more stable resonator, came to the construction in 1980 of the first operational RFQ [8].

The RFQ rapidly became the structure of choice at low energy because it fulfils at the same time three different functions:

i) *focusing* of the particle beam by an electric quadrupole field, particularly valuable at low energy where space charge forces are strong and conventional magnetic quadrupoles are less effective;

ii) *adiabatic bunching* of the beam: starting from the continuous beam produced by the source it creates with minimum beam loss the bunches at the basic RF frequency that are required for acceleration in the subsequent structures;

ii) *acceleration* of the beam from the extraction energy of the source to the minimum required for injection into the following structure.

In modern systems the ion source is followed by a short beam transport required to match transversally the beam coming from the source to the acceptance of the RFQ. Extraction from the ion source (and injection into the RFQ) is usually done at a few tens of keV, achievable with small-sized HV installations. The RFQ then accelerates the beam up to its entrance into the following structure, usually a DTL. Although the RFQ could accelerate the beam to high energy, most of the RF power delivered to the resonator goes to establishing the focusing and bunching field, with the consequence that its acceleration efficiency is very poor. For this reason, RFQs are used only up to a few MeV, in a length of a few metres. Figure 29 shows a photograph of the inside of an RFQ (CERN RFQ1, 202 MHz) and a 3D view of the CERN RFQ for Linac4 (352 MHz), recently commissioned.

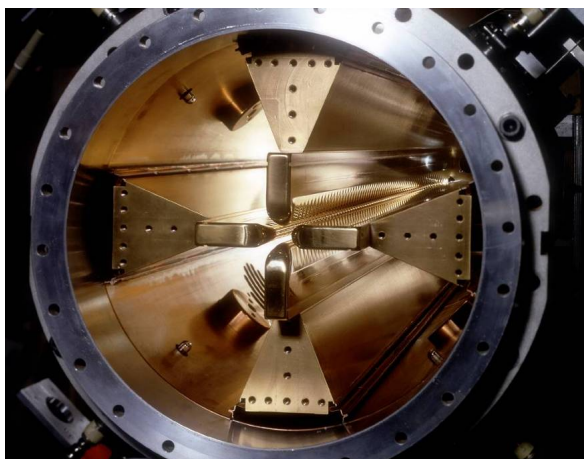
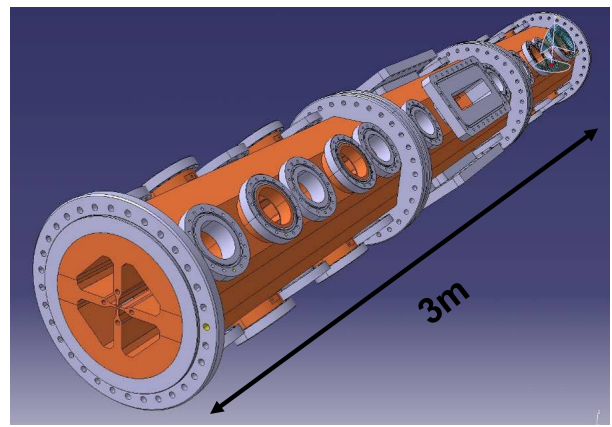

(a)　　　　　　　　　　　　　　　　　　　　(b)

**Fig. 29**: (a) The CERN RFQ1; (b) Linac4 RFQ

The generation of the quadrupole electric field requires four electrodes, visible in Fig. 29(a), which in this particular type of RFQ are called vanes. They are positioned inside a cylindrical tank forming an RF cavity excited in a mode that generates a quadrupole RF voltage between the vane tips (Fig. 30). A particle travelling through the channel formed by the four vanes will see a quadrupole electric field with polarity changing with time, at the period of the RF. Every half RF period the particle will see the polarity of the quadrupole reversed, i.e. it will see an alternating gradient focusing channel, with periodicity corresponding to the distance travelled by the particle during a half RF period, $\beta\lambda/2$. The physics of this electric quadrupole channel is the same as for a magnetic focusing channel where the quadrupole gradient is replaced by the RF voltage and the space periodicity is $\beta\lambda/2$.

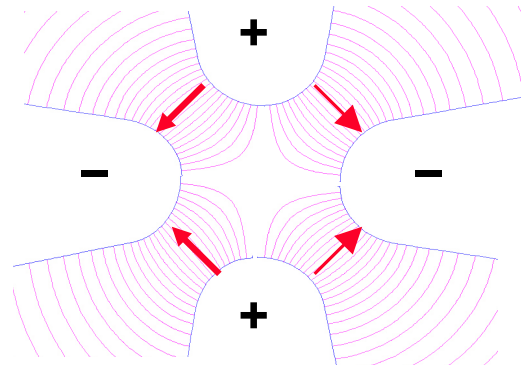

**Fig. 30**: Voltages and electric fields across RFQ vanes

The longitudinal focusing required for bunching and acceleration is provided by a small longitudinal modulation of the vane tips (barely visible in Fig. 29(a). On the tip of the vanes is machined a sinusoidal profile, with period $\beta\lambda$ (Fig 31(a)). The important point, necessary to obtain a longitudinal field component, is that on opposite vanes the peaks and valleys of the modulation correspond, whereas on adjacent (at 90°) vanes the peaks correspond to valleys and vice versa (Fig. 31(b)). The arrows in the scheme for adjacent vanes in Fig. 31(b) represent at a given time the electric field between the two adjacent vanes that have opposite polarity (voltage difference V). On the axis, the electric field vectors can be decomposed in a transverse component, perpendicular to the direction of the beam and, in a small longitudinal component, parallel to the beam direction. The transverse component is constant along the length and represents the focusing field. The longitudinal component instead changes sign (direction) every $\beta\lambda/2$: a particle travelling with velocity $\beta$ will see an accelerating field (or in more general terms, the same RF phase) in every cell, exactly as in a standard π-mode accelerating structure.

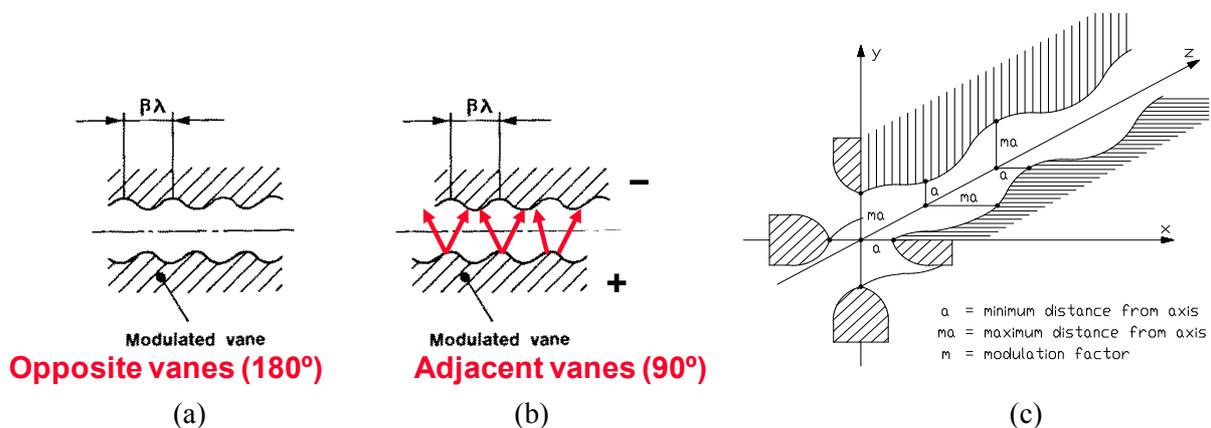

(a)            (b)            (c)

**Fig. 31**: RFQ vanes, field polarity, and modulation parameters

As a result, from the longitudinal point of view an RFQ will be made of a large number of small accelerating cells ($\beta$ being very small at the beginning of the acceleration), with the additional flexibility with respect to conventional structures that it is possible to change from cell to cell: i) the amplitude of the modulation and therefore the intensity of the longitudinal electric field, and ii) the length of the cell and therefore the RF phase seen by the beam in its centre. It is then possible to keep the vanes flat in the initial part of an RFQ (no modulation and only focusing) and after a certain length start ramping up slowly the modulation and the longitudinal field. After the first modulated cells, the particle density will start increasing around the phase at which the RF voltage passes through zero, and the bunch will be slowly formed. Over many cells, the bunching process can be carefully controlled and made adiabatic, with the result of capturing a large fraction of the beam inside the RFQ bucket. When the bunch is formed, the acceleration can start, and the RFQ designer can slowly modify the cell length to bring the centre of the bunch towards the crest of the RF wave. As an example, Fig. 32 shows the evolution of the longitudinal beam emittance (energy vs. phase) in eight cells selected out of the 126 that make up the CERN RFQ2 (90–750 keV): in one of the first modulated cells (top left) the continuous beam coming out of the source sees the first sinusoidal energy modulation; in the following cells (first line) the bunching proceeds until a sufficient density is achieved in the centre (second line) and the acceleration process can start. A few particles are lost in the process, corresponding to the tails visible in the 4th and 5th plot. In the last cells, a bunch is formed, ready for injection into the DTL.

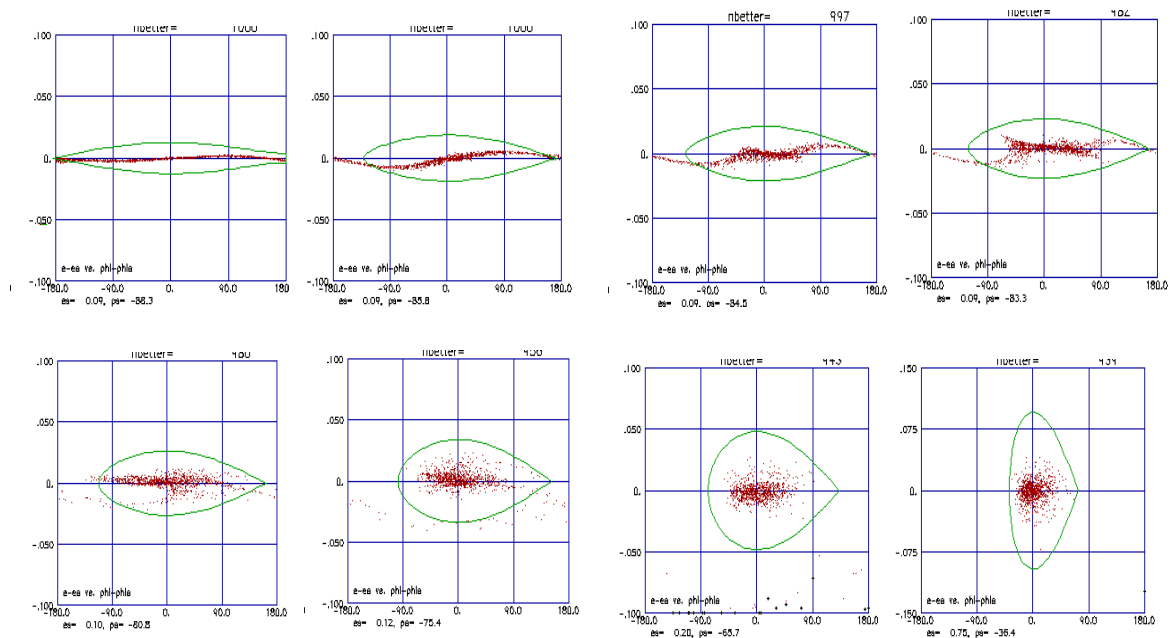

**Fig. 32**: Evolution of longitudinal emittance along the CERN RFQ2, in eight representative cells out of 126

Again, it must be observed that in an RFQ only the last cells are devoted to acceleration; an RFQ is mainly a focusing and bunching device. By correctly defining the parameters of the modulation and the RF voltage, the beam dynamics designer is able to match and transport intense beams, at the same time bunching the beam with minimum particle loss. The drawback is that the beam focusing parameters are frozen forever in the beam modulation and cannot be changed during operation; the RFQ is a 'one-button' machine, where only the RF voltage can be varied during operation. Its design relies completely on the beam transport code, and it is not by coincidence that the development of the RFQs has gone in parallel with the development of modern powerful beam simulation code capable of correctly treating the space charge regime.

## 13  The RFQ RF resonator

From the RF point of view, the problem of building an RFQ consists in creating a time-varying quadrupole-type electric field between four electrodes, keeping the voltage constant (or following a pre-defined law) along its length. To generate this field, the electrodes must be part of an RF resonator; different resonator types can be used, the most-commonly used being the 'four-vane'. It can be considered as a cylindrical resonator excited in TE210 mode, i.e. a quadrupole mode (mode index 2 in the angular polar coordinate) with only transverse electric field components and constant fields along its length (mode index 0 longitudinally). The TE210 mode of the empty cylinder, whose electric and magnetic field symmetry is shown in Fig. 33(a), is transversally 'loaded' by the four vanes that concentrate the electric field on the axis (Fig. 33(b)). The RFQ will result in a cylinder containing the four vanes, which must be connected to the cylinder all along their length.

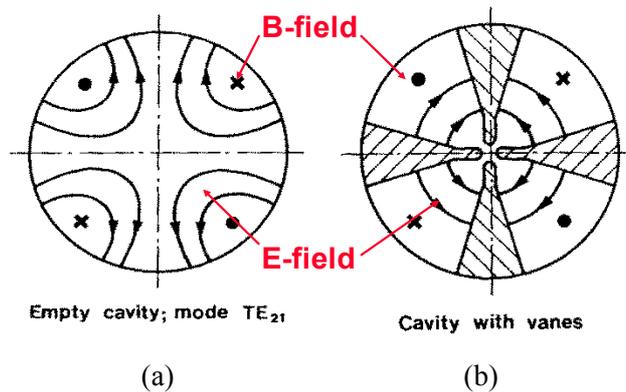

(a)  (b)

**Fig. 33**: Four-vane RFQ. (a) Empty cylinder electric and magnetic field symmetry; (b) four vanes concentrate the electric field on the axis.

The vanes have a two-fold effect on the TE210 mode: on the one hand, they concentrate the quadrupole field on the axis, increasing the RF power efficiency of the structure and the focusing term, and on the other hand they increase the capacitance of this particular mode, decreasing its frequency well below that of the many other modes of the cylindrical resonator. This separation has a positive effect on the stability of the resonator. Unfortunately, the presence of the vanes decreases in the same way the frequency of the TE110 mode, the dipole whose field pattern is shown in Fig. 34. The RFQ resonator will present at a frequency slightly below that of the operating TE210 mode two dipole modes of TE110 type, corresponding to the two orthogonal polarisations of this mode.

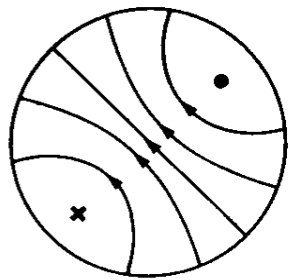

**Fig. 34**: Dipole modes in a cylindrical cavity

The RFQ resonator's high sensitivity to errors coming from the presence of the perturbing modes has to be correctly taken into account in the design, construction, and tuning of the RFQ. To reduce the sensitivity to errors of the RFQ field, alternatives to the four-vane resonator designs have been developed and are in use in many laboratories; of these, the most widely used is the so-called 'four-rod' RFQ (Fig. 35) originally developed by A. Schempp at the IAP of Frankfurt University [9].

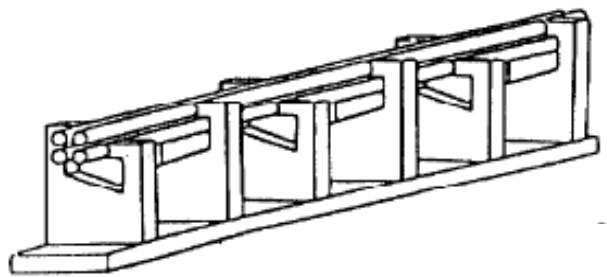

**Fig. 35**: Four-rod RFQ

In this device, the four electrodes are either circular rods with a modulated diameter or small rectangular bars with a modulated profile on one side; they are connected to an array of quarter-wavelength parallel plate transmission lines generating a voltage difference between the two plates (Fig. 36). Opposite pairs of electrodes are connected to the two plates on a line, resulting in a quadrupole voltage being generated between the rods. Several quarter-wavelength cells are used to cover the required RFQ length; their magnetic field couples from each cell to the next, forming a long, single resonator. This 'open' RFQ structure is then placed inside a tank, which forms the vacuum and RF envelope for the structure.

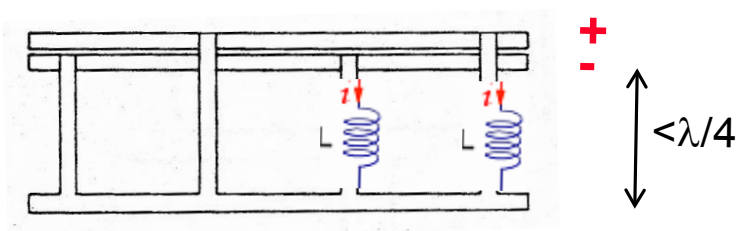

**Fig. 36**: The four-rod RFQ electrical equivalent

The main advantages of the four-rod RFQ are the absence of dipole modes that reduce the sensitivity to mechanical errors and simplifies the tuning, the reduced transverse dimensions as compared to the four-vanes, and the simple and easy-to-access construction. These advantages are particularly evident for the low-frequency RFQs (up to about 100 MHz) used for heavy ions. For the higher frequencies required for protons, from about 200 MHz, the transverse dimensions of the four-rod RFQ become very small and the current and power densities reach high values in some parts of the resonator, in particular at the critical connection between the rods and the supports; cooling can be difficult, in particular for RFQs operating at a high duty cycle, with the risk of excessive deformation of the rods and reduced beam transmission. For these reasons, RFQs operating at frequencies above 200 MHz or at a high duty cycle are usually of the four-vane type.